\documentclass[12pt,preprint]{aastex}
\begin{document}

\title{Correlated Multi-Waveband Variability in the Blazar 3C~279 from 1996 to 2007}

\author{Ritaban Chatterjee\altaffilmark{1}, Svetlana G. Jorstad\altaffilmark{1,2}, Alan P. Marscher\altaffilmark{1}, Haruki Oh\altaffilmark{1,3}, Ian M. McHardy\altaffilmark{4}, Margo F. Aller\altaffilmark{5}, Hugh D. Aller\altaffilmark{5}, Thomas J. Balonek\altaffilmark{6}, H. Richard Miller\altaffilmark{7}, Wesley T. Ryle\altaffilmark{7}, Gino Tosti\altaffilmark{8}, Omar Kurtanidze\altaffilmark{9}, Maria Nikolashvili\altaffilmark{9}, Valeri M. Larionov\altaffilmark{2}, and Vladimir A. Hagen-Thorn\altaffilmark{2}}

\altaffiltext{1}{Institute for Astrophysical Research, Boston University, 725 Commonwealth Avenue, Boston, MA 02215}

\altaffiltext{2}{Astronomical Institute of St. Petersburg State University, Universitetskij Pr. 28, Petrodvorets, 198504 St. Petersburg, Russia}

\altaffiltext{3}{Current address: Department of Physics, University of California, Berkeley, CA 94720-7300}

\altaffiltext{4}{Department of Physics and Astronomy, University of Southampton, Southampton, SO17 1BJ, United Kingdom}

\altaffiltext{5}{Astronomy Department, University of Michigan, 830 Dennison, 501 East University Street, Ann Arbor, Michigan 48109-1090}

\altaffiltext{6}{Department of Physics and Astronomy, Colgate University, 13 Oak Drive, Hamilton, NY 13346}

\altaffiltext{7}{Department of Physics and Astronomy, Georgia State University, Atlanta, GA 30303}

\altaffiltext{8}{Department of Physics, University of Perugia, Via A. Pascoli, 06123 Perugia, Italy}

\altaffiltext{9}{Abastumani Astrophysical Observatory, Mt. Kanobili, Abastumani, Georgia}

\begin{abstract}
We present the results of extensive multi-waveband monitoring of the blazar 3C~279 between 1996 and 2007 at X-ray energies (2-10 keV), optical R band, and 14.5 GHz, as well as imaging with the Very Long Baseline Array (VLBA) at 43 GHz. In all bands the power spectral density corresponds to ``red noise'' that can be fit by a single power law over the sampled time scales. Variations in flux at all three wavebands are significantly correlated. The time delay between high and low frequency bands changes substantially on time scales of years. A major multi-frequency flare in 2001 coincided with a swing of the jet toward a more southerly direction, and in general the X-ray flux is modulated by changes in the position angle of the jet near the core. The flux density in the core at 43 GHz---increases in which indicate the appearance of new superluminal knots---is significantly correlated with the X-ray flux.

We decompose the X-ray and optical light curves into individual flares, finding that X-ray leads optical variations (XO) in 6 flares, the reverse occurs in 3 flares (OX), and there is essentially zero lag in 4 flares. Upon comparing theoretical expectations with the data, we conclude that (1) XO flares can be explained by gradual acceleration of radiating electrons to the highest energies; (2) OX flares can result from either light-travel delays of the seed photons (synchrotron self-Compton scattering) or gradients in maximum electron energy behind shock fronts; and (3) events with similar X-ray and optical radiative energy output originate well upstream of the 43 GHz core, while those in which the optical radiative output dominates occur at or downstream of the core.
\end{abstract}

\section{Introduction}   
Blazars form a subclass of active galactic nuclei (AGN) characterized by violent variability on time scales from hours to years across the electromagnetic spectrum. It is commonly thought that at radio, infrared, and optical frequencies the variable emission of blazars originates in relativistic jets and is synchrotron in nature \citep{imp88,mar98}. The X-ray emission is consistent with inverse Compton (IC) scattering of these synchrotron photons, although seed photons from outside the jet cannot be excluded \citep{mau96,rom97,cop99,bla00,sik01,chi02,arb05}. The models may be distinguished by measuring time lags between the seed-photon and Compton-scattered flux variations. Comparison of the amplitudes and times of peak flux of flares at different wavebands is another important diagnostic. For this reason, long-term multi-frequency monitoring programs are crucially important for establishing a detailed model of blazar activity and for constraining the physics of relativistic jets. Here we report on such a program that has followed the variations in emission of the blazar 3C~279 with closely-spaced observations over a time span of $\sim$10 years.

The quasar 3C~279 \citep[redshift=0.538;][]{bur65} is one of the most prominent blazars owing to its high optical polarization and variability of flux across the electromagnetic spectrum. Very long baseline interferometry (VLBI) reveals a one-sided radio jet featuring bright knots (components) that are ``ejected'' from a bright, presumably stationary ``core'' \citep{jor05}. The measured apparent speeds of the knots observed in the past range from $4c$ to $16c$ \citep{jor04},  superluminal motion that results from relativistic bulk velocities and a small angle between the jet axis and line of sight. Relativistic Doppler boosting of the radiation increases the apparent luminosity to $\sim 10^4$ times the value in the rest frame of the emitting plasma.

Characterization of the light curves of 3C~279 includes the power spectral density (PSD) of the variability at all different wavebands.The PSD corresponds to the power in the variability of emission as a function of time scale.
\citet{law87} and \citet{mch87} have found that the PSDs of many Seyfert galaxies, in which the continuum emission comes mainly from the central engine, are simple ``red noise'' power laws, with slopes between $-1$ and $-2$. More recent studies indicate that some Seyfert galaxies have X-ray PSDs that are fit better by a broken power law, with a steeper slope above the break frequency \citep{utt02, mch04, mar03, ede99, pou01}. This property of Seyferts is similar to that of Galactic black hole X-ray binaries (BHXRB), whose X-ray PSDs contain one or more breaks \citep{bel90,now99}. However, in blazars most of the X-rays are likely produced in the jets rather than near the central engine as in BHXRBs and Seyferts. It is unclear {\it a priori} what the shape of the PSD of nonthermal emission from the jet should be, a question that we answer with the dataset presented here.

The raw PSD calculated from a light curve combines two aspects of the dataset: (1) the intrinsic variation of the object and (2) the effects of the temporal sampling pattern of the observations. In order to remove the latter, we apply a Monte-Carlo type algorithm based on the ``Power Spectrum Response Method'' (PSRESP) of \citet{utt02} to determine the intrinsic PSD (and its associated uncertainties) of the light curve of 3C~279 at each of three wavebands. Similar complications affect the determination of correlations and time lags of variable emission at different wavebands. Uneven sampling, as invariably occurs, can cause the correlation coefficients to be artificially low. In addition, the time lags can vary across the years owing to physical changes in the source. In light of these issues, we use simulated light curves, based on the underlying PSD, to estimate the significance of the derived correlation coefficients.

In {\S}2 we present the observations and data reduction procedures, while in {\S}3 we describe the power spectral analysis and its results. In {\S}4 we cross-correlate the light curves to determine the relationship of the emission at different wavebands. Finally, in {\S}5 we discuss and interpret the results, focusing on the implications regarding the location of the nonthermal radiation at different frequencies, as well as the physical processes in the jet that govern the development of flux outbursts in blazars.

\section{Observations and Data Analysis}
\subsection{Monitoring of the X-ray, Optical, and Radio Flux Density}
Table~\ref{data} summarizes the intervals of monitoring at different frequencies for each of the three wavebands in our program. We term the entire light curve ``monitor data''; shorter segments of more intense monitoring are described below.

The X-ray light curves are based on observations of 3C~279 with the Rossi X-ray Timing Explorer (RXTE) from 1996 to 2007. We observed 3C~279 in 1222 separate pointings with the RXTE, with a typical spacing of 2-3 days. The exposure time varied, with longer on-source times---typically 1-2 ks---after 1999 as the number of fully functional detectors decreased, and shorter times at earlier epochs. For each exposure, we used routines from the X-ray data analysis software FTOOLS and XSPEC to calculate and subtract an X-ray background model from the data and to model the source spectrum from 2.4 to 10 keV as a power law with low-energy photoelectric absorption by intervening gas in our Galaxy. For the latter, we used a hydrogen column density of $8\times 10^{20}$ atoms cm$^{-2}$. There is a $\sim$ 1-year gap in 2000 and annual 8-week intervals when the quasar is too close to the Sun's celestial position to observe. A small part of the X-ray data is given in Table~\ref{xdatatable}. The whole dataset will be available in the electronic version of ApJ.

\begin{figure}
\epsscale{.80}
\plotone{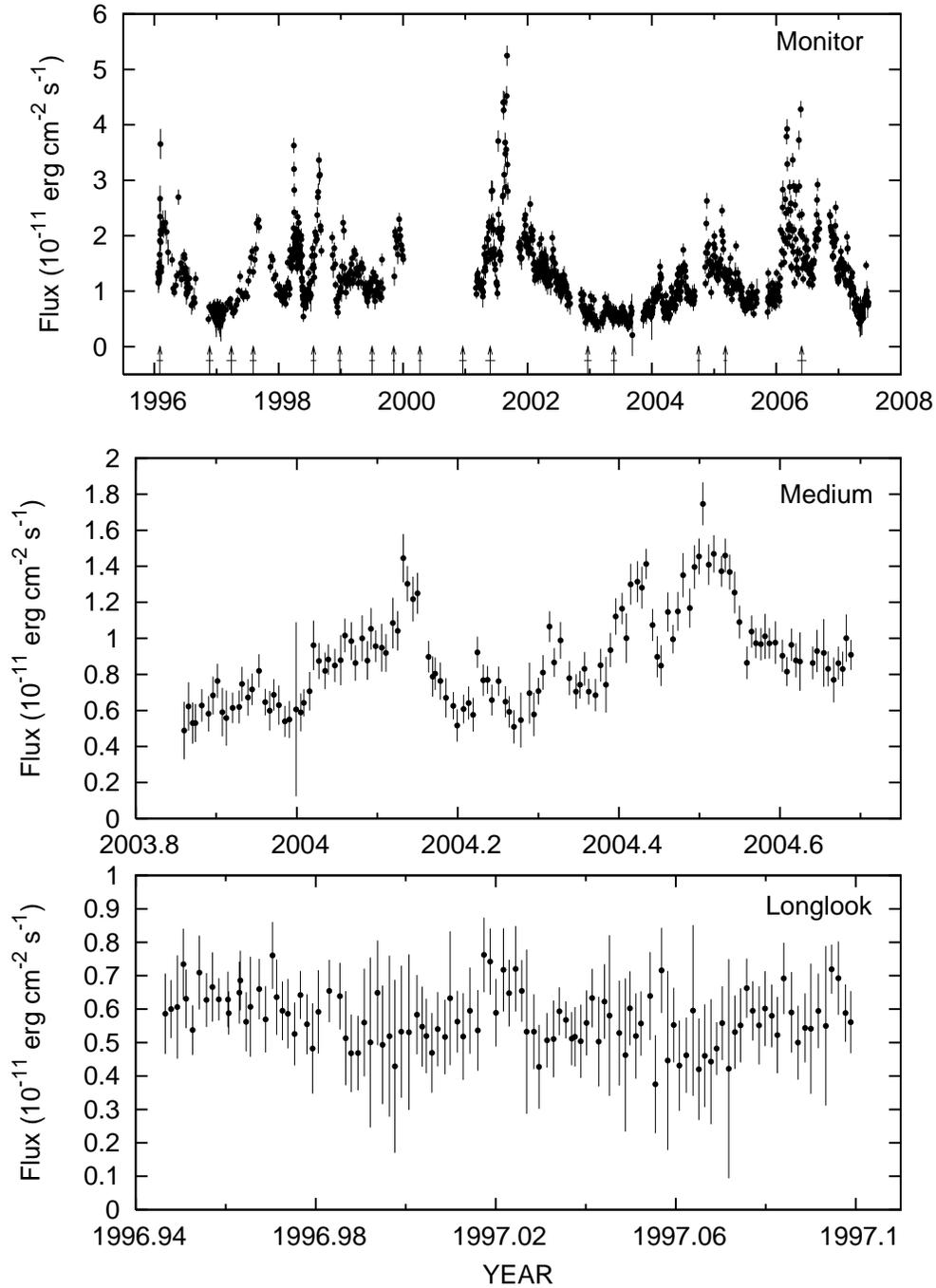}
\caption{X-ray (2-10 keV) data on different time-scales. In the upper panel, the arrows show the times of superluminal ejections and the line segments perpendicular to the arrows show the uncertainties in the times of ejection.}
\label{xdata}
\end{figure}

In 1996 December and 1997 January we obtained, on average, two measurements per day for almost two months. We refer to these observations as the ``longlook" data. Between 2003 November and 2004 September, we obtained 127 measurements over 300 days (the ``medium" data). Figure~\ref{xdata} presents these three datasets. The X-ray spectral index $\alpha_x$, defined by $f_{\rm x} \propto \nu^{\alpha_{\rm x}}$, where $f_{\rm x}$ is the X-ray flux density and $\nu$ is the frequency, has an average value of $-0.8$ with a standard deviation of 0.2 over the $\sim 10$ years of observation, and remained negative throughout.

\begin{figure}
\epsscale{.80}
\plotone{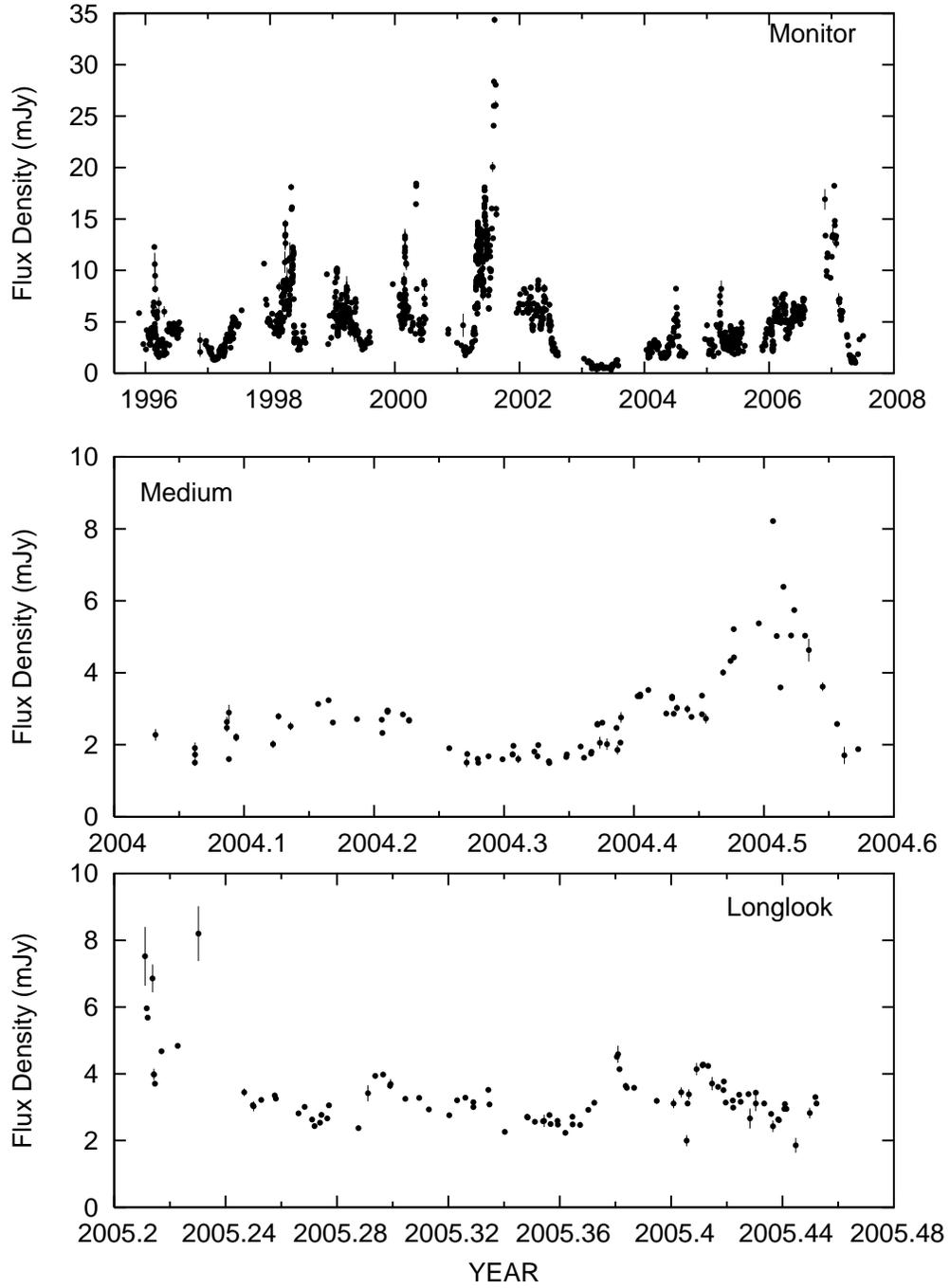}
\caption{Optical (R-band) data on different time-scales.}
\label{opdata}
\end{figure}

We monitored 3C~279 in the optical $R$ band over the same time span as the X-ray observations. The majority of the measurements between 1996 and 2002 are from the 0.3 m telescope of the Foggy Bottom Observatory, Colgate University, in Hamilton, New York. Between 2004 and 2007, the data are from the 2 m Liverpool Telescope (LT) at La Palma, Canary Islands, Spain, supplemented by observations at the 1.8 m Perkins Telescope of Lowell Observatory, Flagstaff, Arizona, 0.4 m telescope of the University of Perugia Observatory, Italy, 0.7 m telescope at the Crimean Astrophysical Observatory, Ukraine, the 0.6 m SMARTS consortium telescope at the Cerro Tololo Inter-American Observatory, Chile, and the 0.7 m Meniscus Telescope of Abastumani Astrophysical Observatory in Abastumani, Republic of Georgia. We checked the data for consistency using overlapping measurements from different telescopes, and applied corrections, if necessary, to adjust to the LT system. We processed the data from the LT, Perkins Telescope, Crimean Astrophysical Observatory, and Abastumani Astrophysical Observatory in the same manner, using comparison stars 2, 7, and 9 from \citet{gon01} to determine the magnitudes in R band. The frequency of optical measurements over the $\sim 10$-year span presented here is, on average, 2-3 observations per week. Over a three-month period between 2005 March and June, we obtained about 100 data points, i.e., almost one per day (``longlook" data). Another subset (``medium'') contains $\sim 100$ points over 200 days between 2004 January and July. Figure~\ref{opdata} displays these segments along with the entire 10-year light curve. A small part of the optical data is given in Table~\ref{opdatatable}. The whole dataset will be available in the electronic version of ApJ.

\begin{figure}
\epsscale{.80}
\plotone{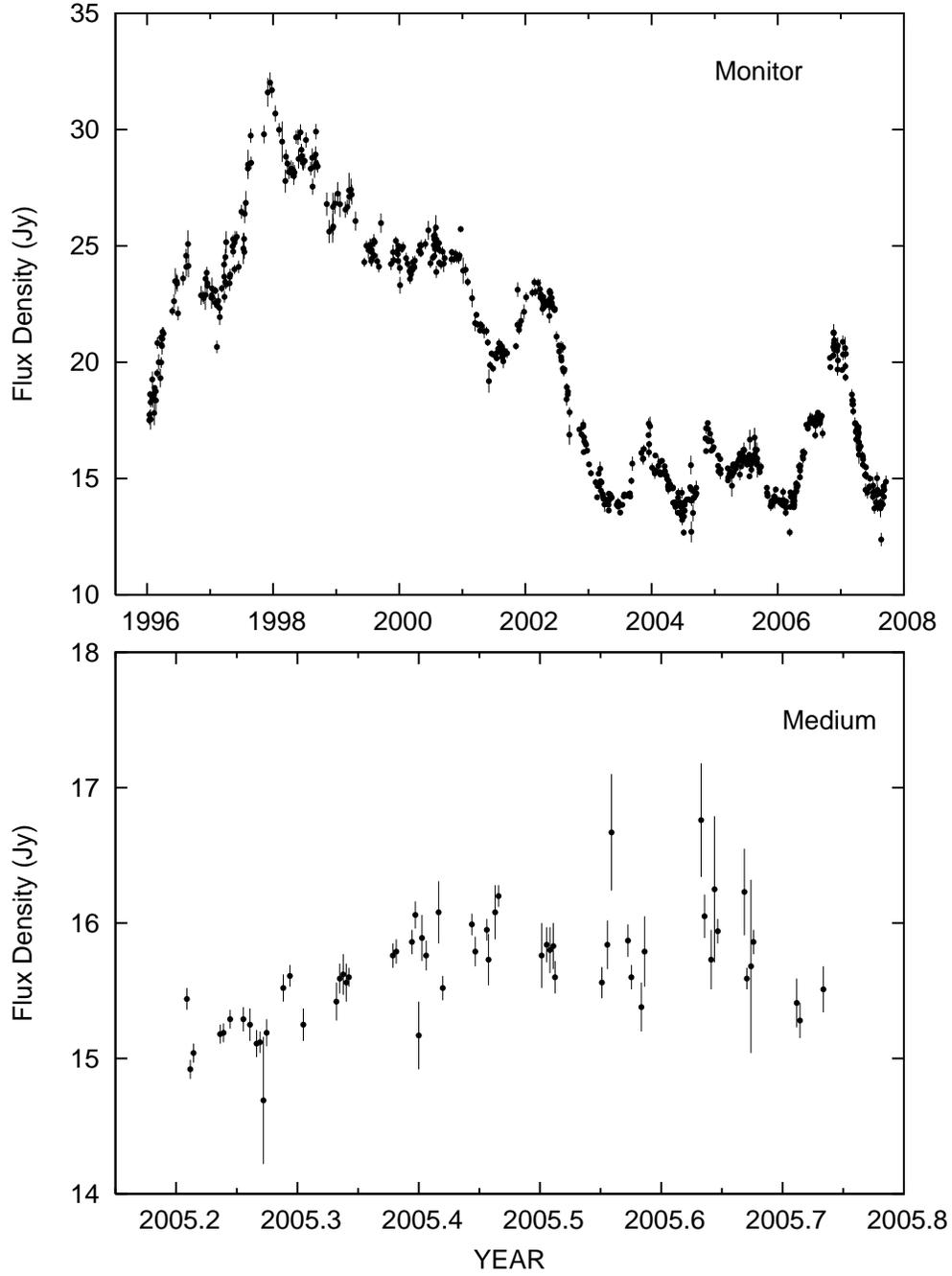}
\caption{Radio (14.5 GHz) data on different time-scales.}
\label{raddata}
\end{figure}

We have compiled a 14.5 GHz light curve (Fig.~\ref{raddata}) with data from the 26 m antenna of the University of Michigan Radio Astronomy Observatory. Details of the calibration and analysis techniques are described in \citet{all85}. The flux scale is set by observations of Cassiopeia A \citep{baa77}. The sampling frequency was usually of order once per week. An exception is a span of about 190 days between 2005 March and September when we obtained 60 measurements, averaging one observation every $\sim 3$ days (``medium" data). A small part of the radio data is given in Table~\ref{raddatatable}. The whole dataset will be available in the electronic version of ApJ.

\subsection{Ultra-high Resolution Images with the Very Long Baseline Array}
\begin{figure}
\epsscale{.5}
\plotone{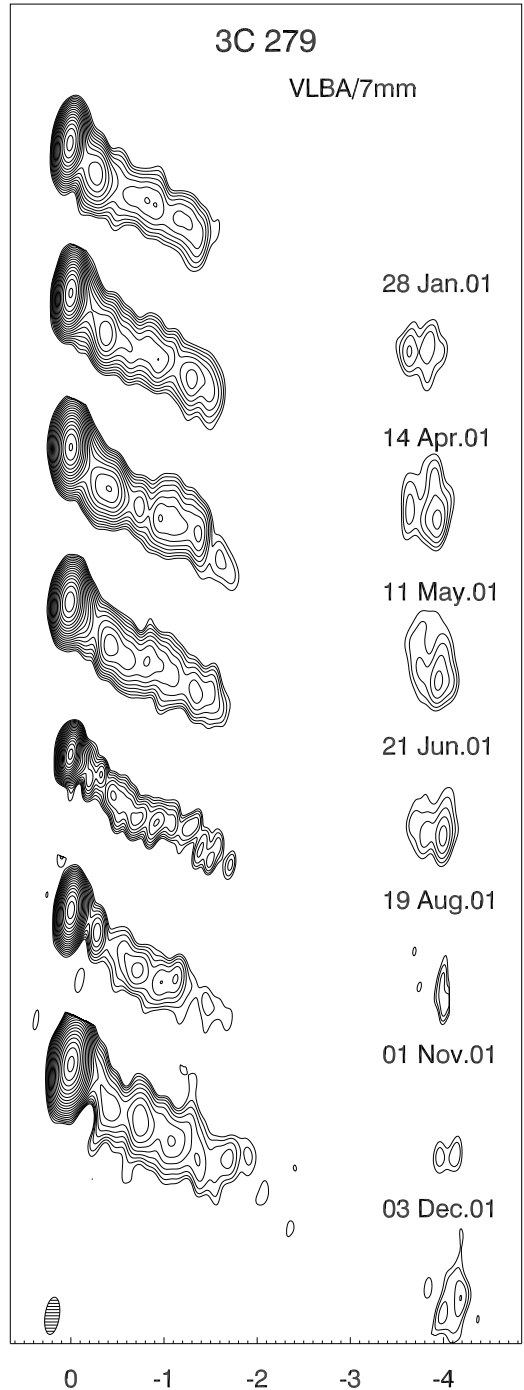}
\caption{Sequence of VLBA images at 7 mm during 2001. The images are convolved with the beam of the size 0.38$\times$0.14 mas at PA = $-9\degr$. The global peak over all maps is 17.24 Jy/Beam. The contour levels are 0.25, 0.354, 0.5, 0.707, ..., 90.51 \% of the global peak. The angular scale given at the bottom is in milliarcseconds (mas).}
\label{vlbaimage1}
\end{figure}

\begin{figure}
\epsscale{.5}
\plotone{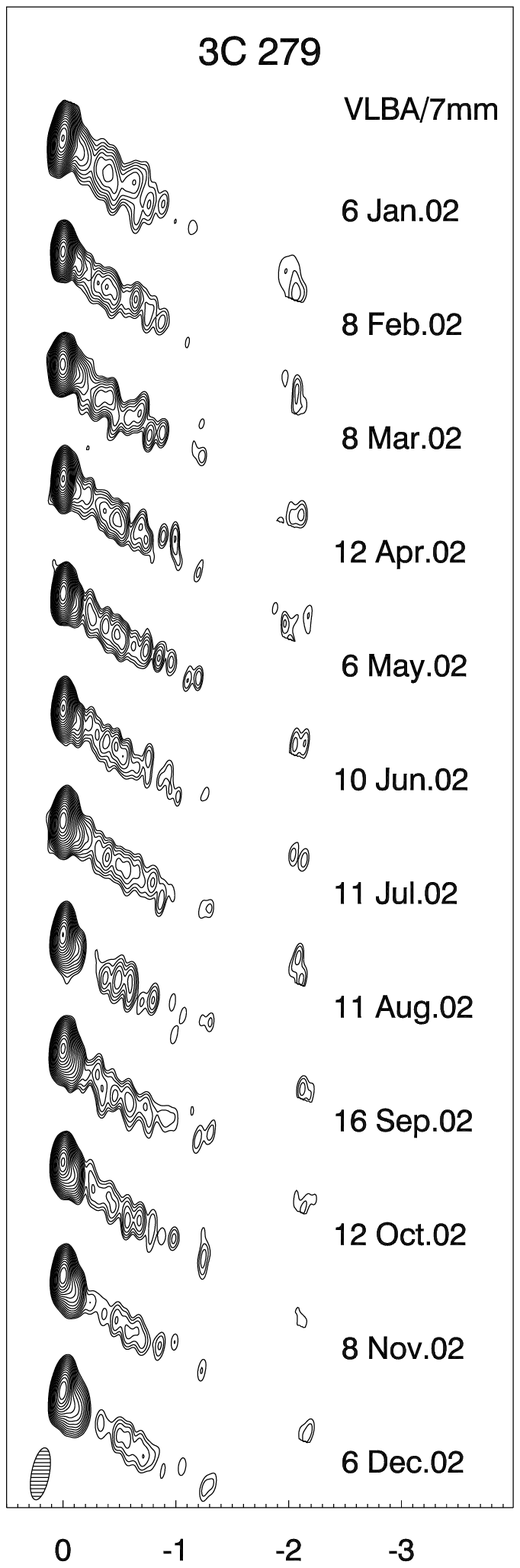}
\caption{Sequence of VLBA images at 7 mm during 2002. For details see caption of Figure 4.}
\label{vlbaimage2}
\end{figure}

\begin{figure}
\epsscale{.5}
\plotone{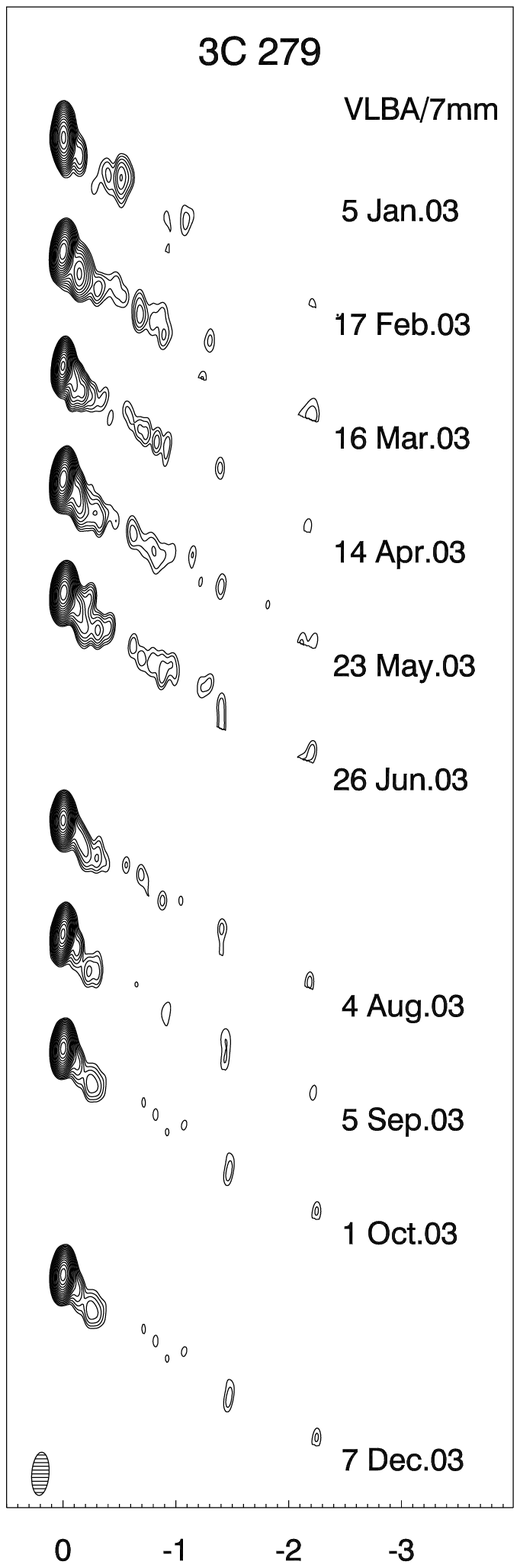}
\caption{Sequence of VLBA images at 7 mm during 2003. For details see caption of Figure 4.}
\label{vlbaimage3}
\end{figure}

\begin{figure}
\epsscale{.5}
\plotone{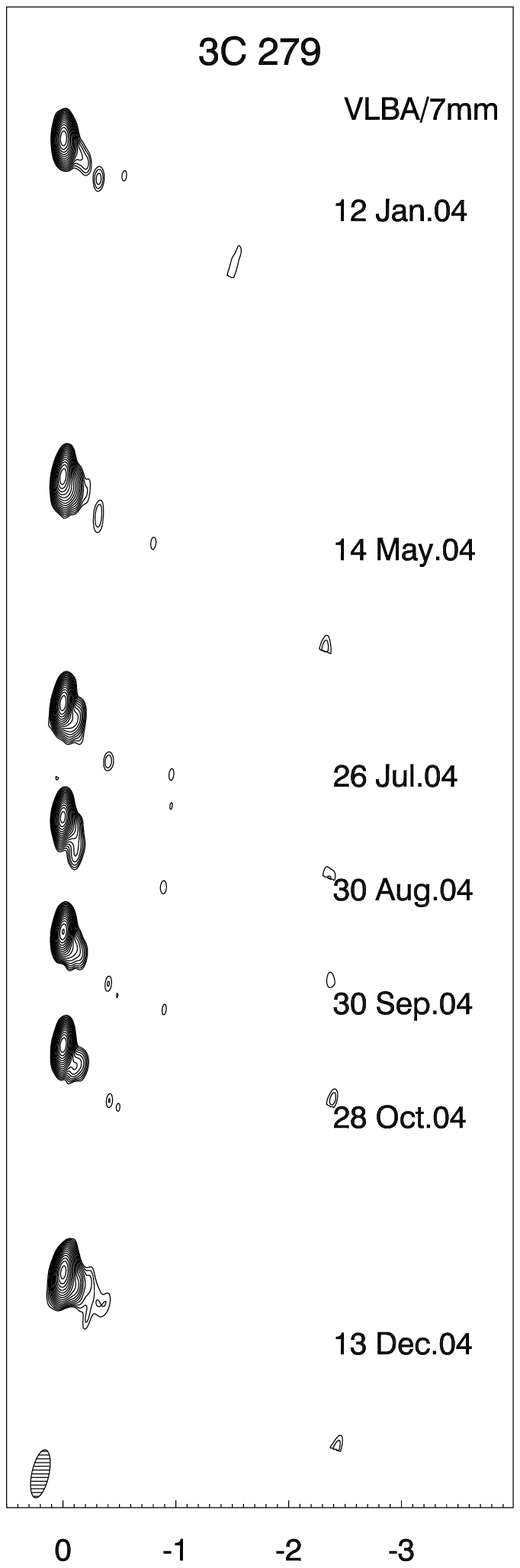}
\caption{Sequence of VLBA images at 7 mm during 2004. For details see caption of Figure 4.}
\label{vlbaimage4}
\end{figure}

\begin{figure}
\epsscale{.5}
\plotone{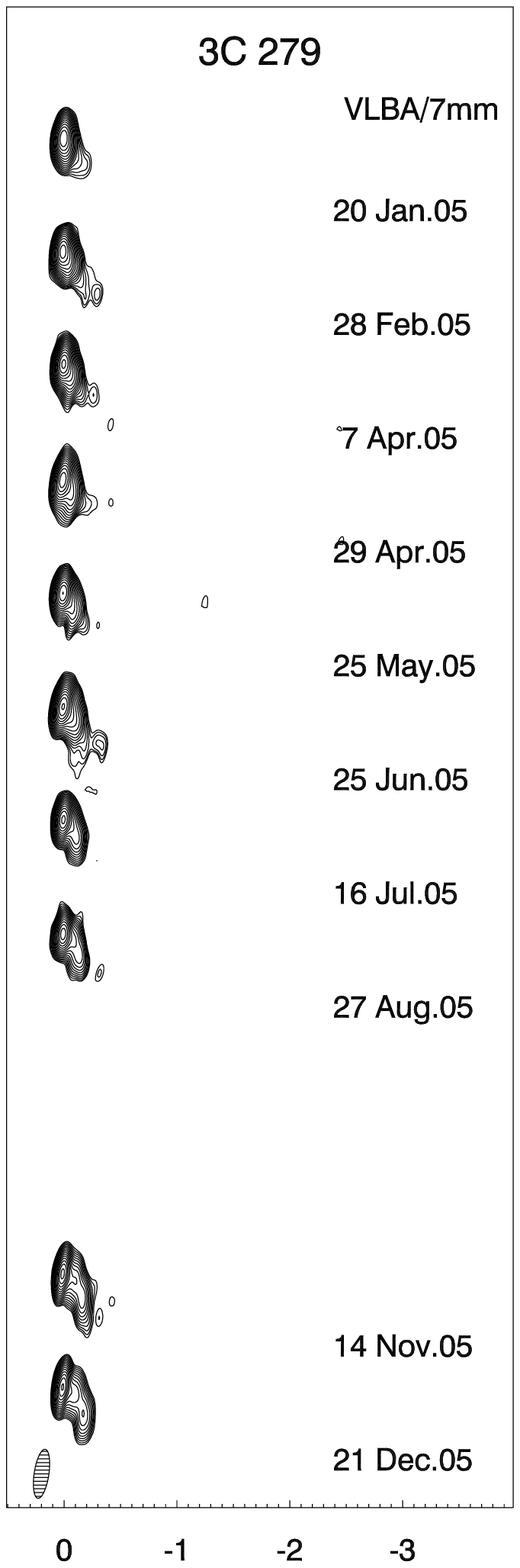}
\caption{Sequence of VLBA images at 7 mm during 2005. For details see caption of Figure 4.}
\label{vlbaimage5}
\end{figure}

\begin{figure}
\epsscale{.5}
\plotone{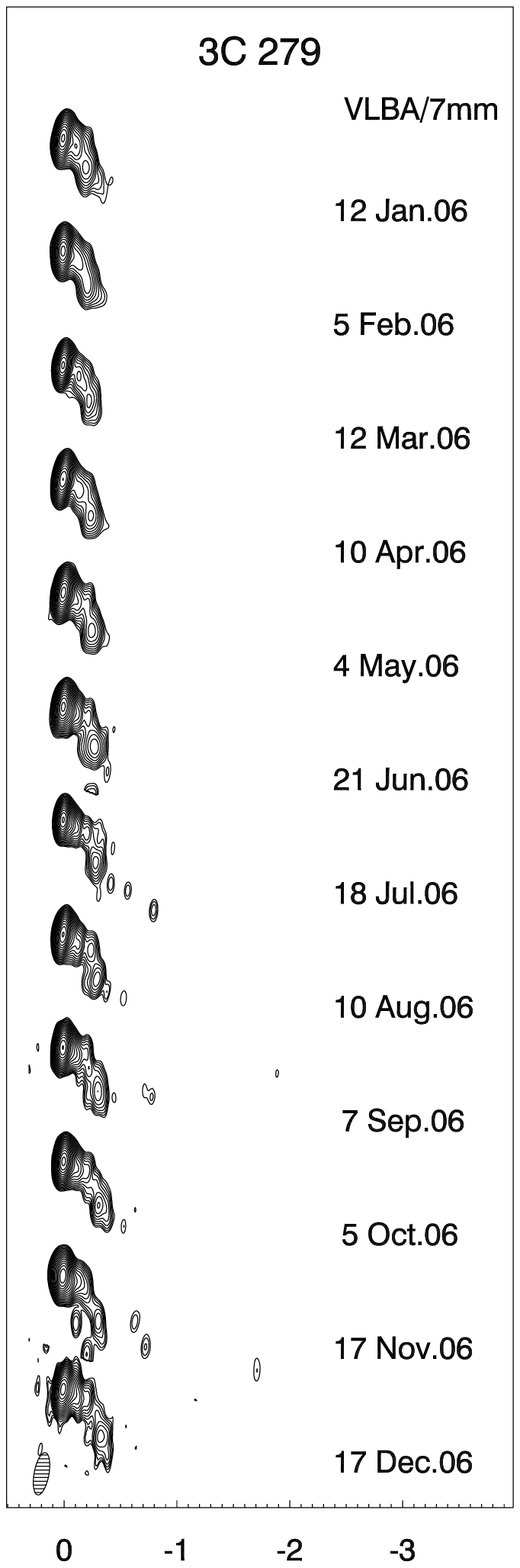}
\caption{Sequence of VLBA images at 7 mm during 2006. For details see caption of Figure 4.}
\label{vlbaimage6}
\end{figure}

\begin{figure}
\epsscale{.99}
\plotone{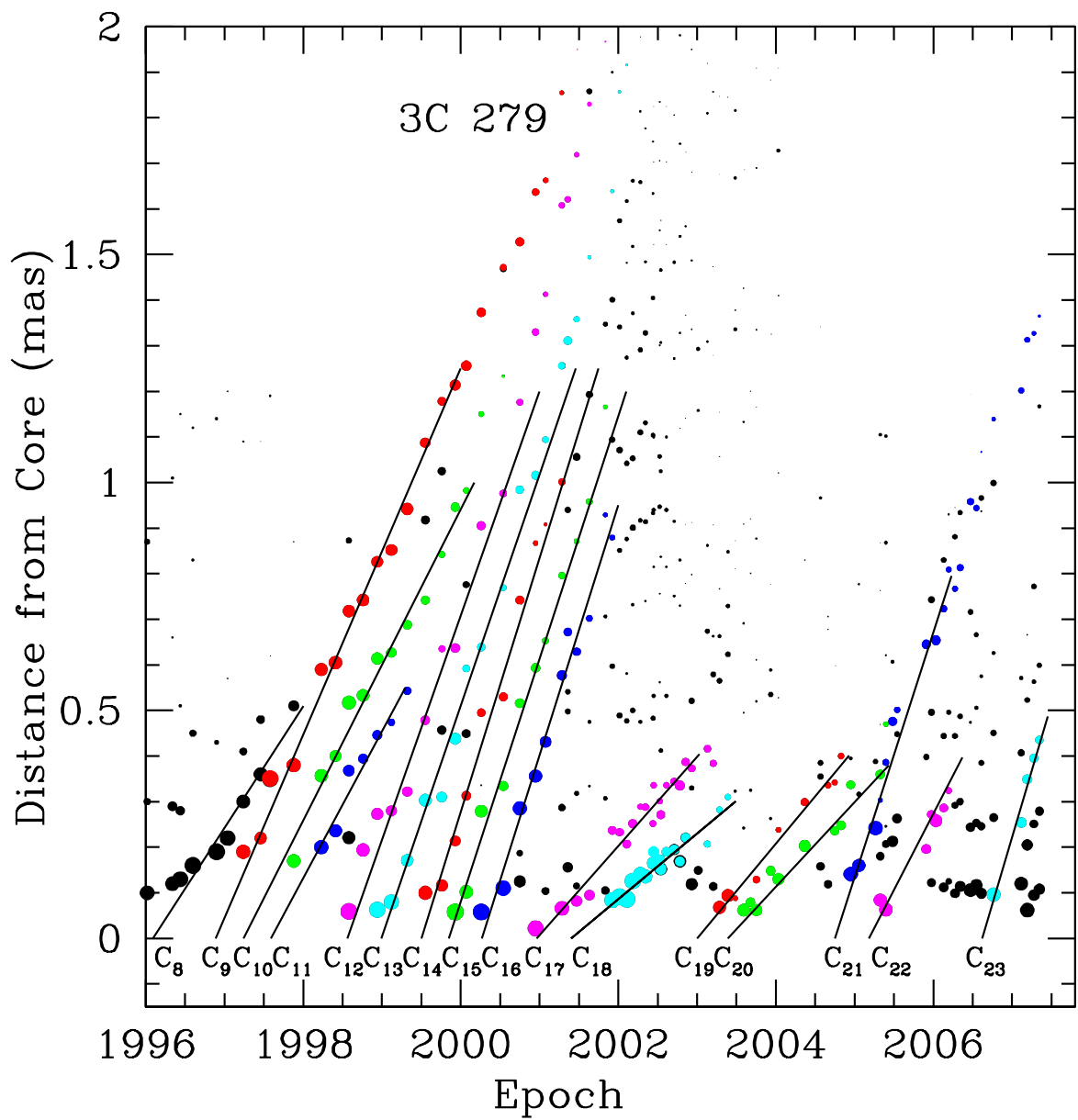}
\caption{Angular separation from the core vs. epoch of all knots brighter than 100 mJy within 2.0 mas of the core. The black lines indicate the motion of each knot (denoted by a given color of data points) listed in Table 2. A knot is identified through continuity of the trajectory from one epoch to the next. The diameter of each symbol is proportional to the logarithm of the flux density of the knot, as determined by model fitting of the VLBA data.}
\label{distepoch}
\end{figure}

Starting in 2001 May, we observed 3C~279 with the Very Long Baseline Array (VLBA) at roughly monthly intervals, with some gaps of 2-4 months. The sequence of images from these data (Fig.~\ref{vlbaimage1} to Fig.~\ref{vlbaimage6}) provides a dynamic view of the jet at an angular resolution $\sim 0.1$ milliarcseconds (mas). We processed the data in the same manner as described in \citet{jor05}. For epochs from 1995 to 2001, we use the images and results of \citet{lis98}, \citet{weh01}, and \citet{jor01,jor05}. We model the brightness distribution at each epoch with multiple circular Gaussian components using the task MODELFIT of the software package DIFMAP \citep{she97}. At each of the 80 epochs of VLBA observation since 1996, this represents the jet emission downstream of the core by a sequence of knots (also referred to as ``components''), each characterized by its flux density, FWHM diameter, and position relative to the core.
Figure~\ref{distepoch} plots the distance vs. epoch for all components brighter than 100 mJy within 2.0 mas of the core. We use the position vs. time data to determine the projected direction on the sky of the inner jet, as well as the apparent speeds and birth dates of new superluminal knots. 

\begin{figure}
\epsscale{.80}
\plotone{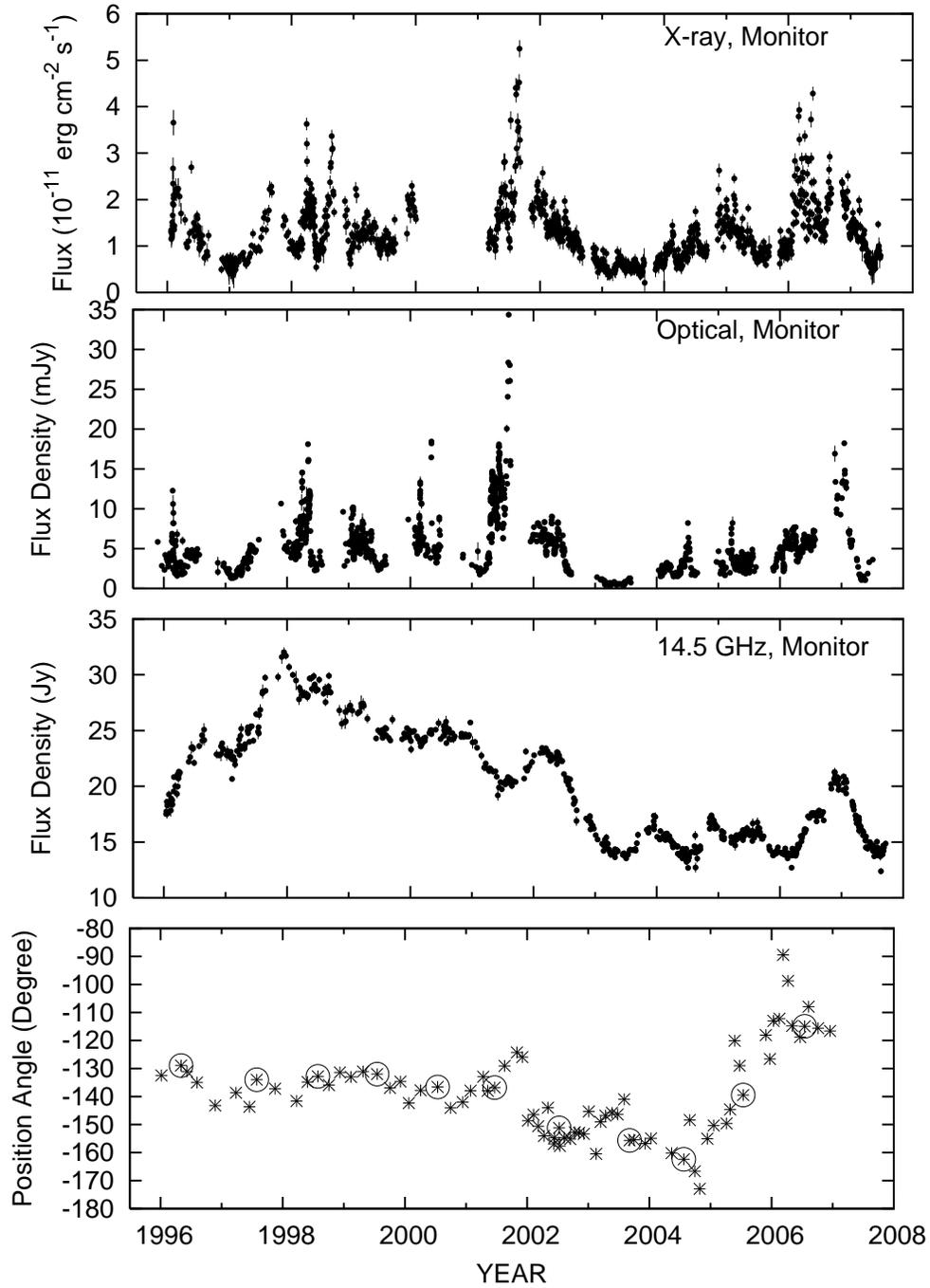}
\caption{Variation of X-ray flux, optical flux, radio flux and position angle of the jet from 1996 to 2008. The circled data points in the bottom panel are the epochs shown in Fig.~\ref{vlba12}.}
\label{xoppa}
\end{figure}

\begin{figure}
\epsscale{.45}
\plotone{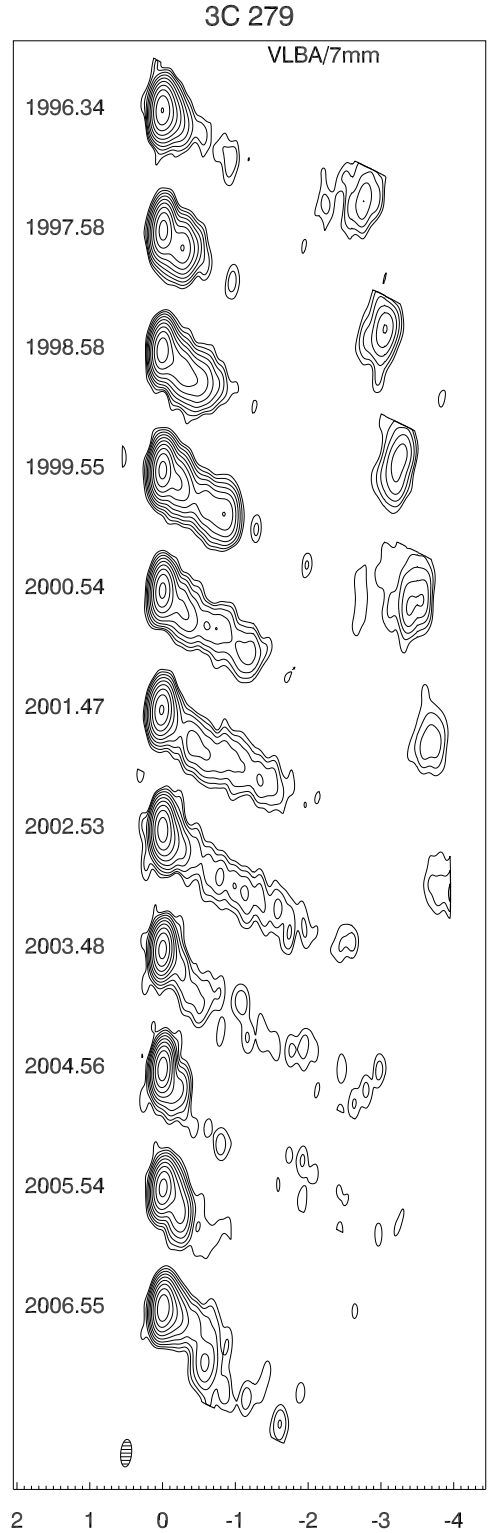}
\caption{VLBA images at one epoch during each year of 11-year monitoring. The images are convolved with the beam of the size 0.38 $\times$0.14 mas at PA = $-$9$\degr$. The map peak is 17.0 Jy/Beam. The contour levels are 0.15, 0.3, 0.6, ...,76.8 \% of the peak. The angular scale given at the bottom is in milliarcseconds (mas). The circled points in Fig.~\ref{xoppa} (bottom panel) correspond to these images.}
\label{vlba12}
\end{figure}

We define the inner-jet position angle (PA) $\theta_{\rm jet}$ with respect to the core as that of the brightest component within 0.1-0.3 mas of the core. As seen in Figure~\ref{xoppa}, $\theta_{\rm jet}$ changes significantly ($\sim 80\degr$) over the 11 years of VLBA monitoring.
Figure~\ref{vlba12} displays a sampling of the VLBA images at epochs corresponding to the circled points in the lower panel of Figure~\ref{xoppa}. 

\begin{figure}
\rotate
\epsscale{.80}
\plotone{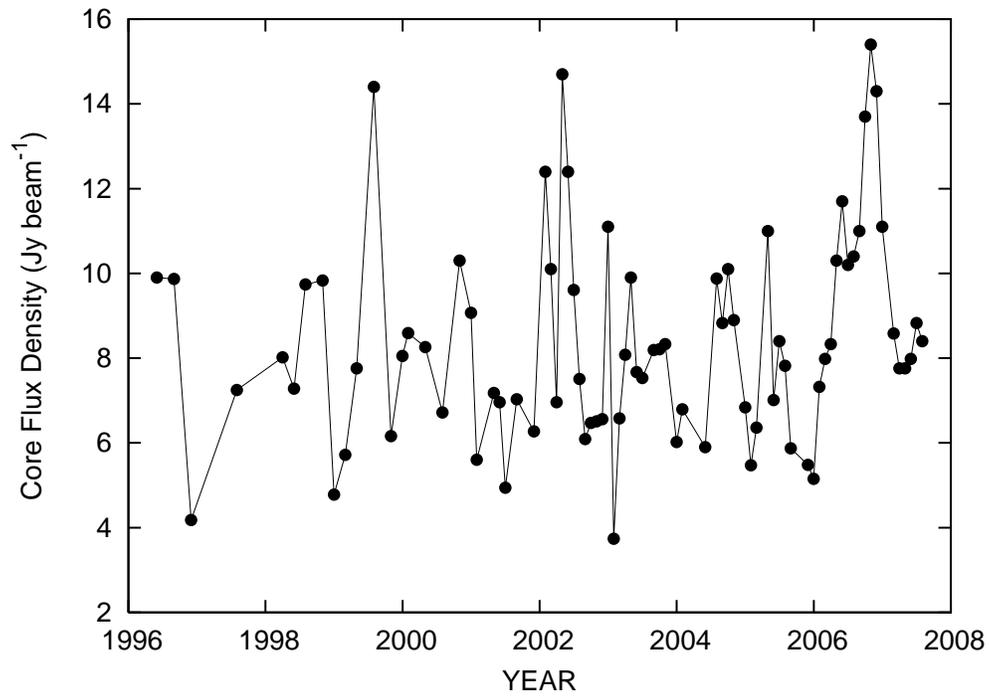}
\caption{Light curve of the VLBA core region at 43 GHz. The jagged line through the data points is drawn solely to aid the eye to follow the variability. Statistical and systematic uncertainty in each measurement is difficult to determine accurately, but is typically 10-20\%.}
\label{corelc}
\end{figure}

We determine the apparent speed $\beta_{app}$ of the moving components using the same procedure as defined in \citet{jor05}. The ejection time $T_0$ is the extrapolated epoch of coincidence of a moving knot with the position of the (presumed stationary) core in the VLBA images. In order to obtain the most accurate values of $T_0$, given that non-ballistic motions may occur \citep{jor04,jor05}, we use only those epochs when a component is within 1 mas of the core, inside of which we assume its motion to be ballistic. $\theta_{\rm jet}$, $T_0$, and $\beta_{app}$ between 1996 and 2007 are shown in Table~\ref{ejecflare}. 
As part of the modeling of the images, we have measured the flux density of the unresolved core in all the images, and display the resulting light curve in Figure~\ref{corelc}.

\section{Power Spectral Analysis and Results}
For all three wavebands, we used the monitor, medium, and longlook data to calculate the PSD at the low, intermediate, and high frequencies of variation, respectively. Since we do not have any longlook data at 14.5 GHz, we determine the radio PSD up to the highest variational frequency that can be achieved with the medium data.

\subsection{Calculation of the PSD of the Observed Light Curve}
We follow \citet{utt02} to calculate the PSD of a discretely sampled light curve $f(t_i)$ of length $N$ points using the formula \\
\begin{equation}
|F_{N}(\nu)|^{2} = \left[\sum_{i=1}^{N}f(t_{i})\cos(2\pi\nu t_{i})\right]^{2} + \left[\sum_{i=1}^{N}f(t_{i})\sin(2\pi\nu t_{i})\right]^{2}.\\
\end{equation}
This is the square of the modulus of the discrete Fourier transform of the (mean subtracted) light curve, calculated for evenly spaced frequencies between $\nu_{\rm min}$ and $\nu_{\rm max}$, i.e., $\nu_{\rm min}$, 2$\nu_{\rm min}$, . . ., $\nu_{\rm max}$.  Here, $\nu_{\rm min}$=1/$T$ ($T$ is the total duration of the light curve, $t_{N}-t_{1}$) and $\nu_{\rm max}$=$N$/2$T$ equals the Nyquist frequency $\nu_{\rm Nyq}$. We use the following normalization to calculate the final PSD:\\ 
\begin{equation}
P(\nu) = \frac{2T}{\mu^{2}N^{2}}|F_{N}(\nu)|^{2},  
\end{equation}
where $\mu$ is the average flux density over the light curve.

We bin the data in time intervals $\Delta T$ ranging from 0.5 to 25 days, as listed in Table~\ref{lcprop}, averaging all data points within each bin to calculate the flux. For short gaps in the time coverage, we fill empty bins through linear interpolation of the adjacent bins in order to avoid gaps that would distort the PSD. We account for the effects of longer gaps, such as sun-avoidance intervals and the absence of X-ray data in 2000, by inserting in each of the simulated light curves the same long gaps as occur in the actual data.

\subsection{PSD Results: Presence/absence of a Break}
We use a variant of PSRESP \citep{utt02} to determine the intrinsic PSD of each light curve. The method is described in the Appendix. PSRESP gives both the best-fit PSD model and a ``success fraction'' that indicates the goodness of fit of the model. 

\begin{figure}
\epsscale{.80}
\plotone{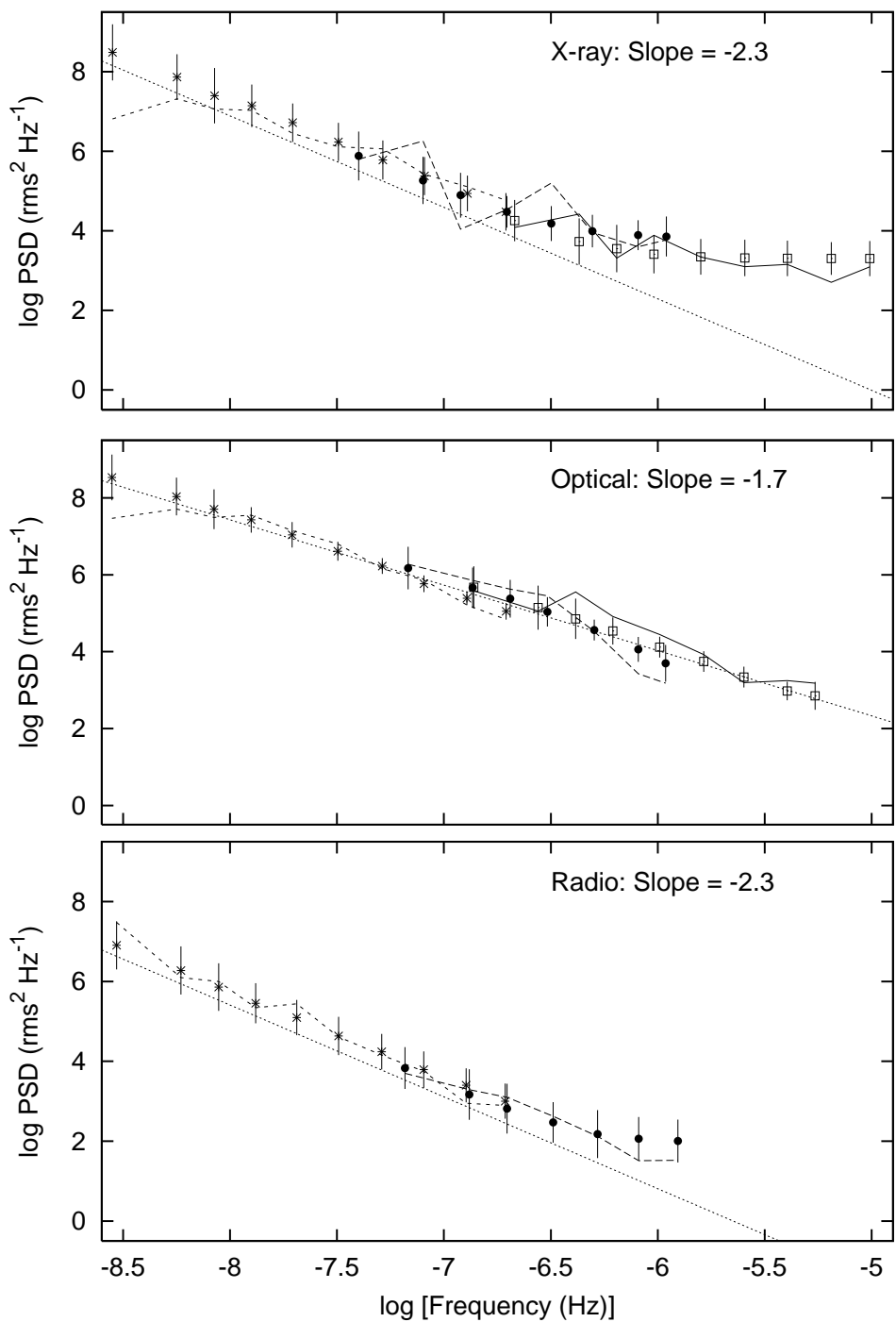}
\caption{Result of application of the PSRESP method to the light curves. PSD of the observed data at high, medium and low frequency range are given by the solid, dashed and dotted jagged lines respectively while the underlying power-law model is given by the dotted straight line. Points with error bars (open squares, solid circles and asterisks for high, medium and low frequency range respectively) correspond to the mean value of the PSD simulated from the underlying power-law model (see text). The errorbars are the standard deviation of the distribution of simulated PSDs. The broadband PSD in all three wavelengths can be described by a simple power law.}
\label{psd}
\end{figure}

\begin{figure}
\epsscale{.80}
\plotone{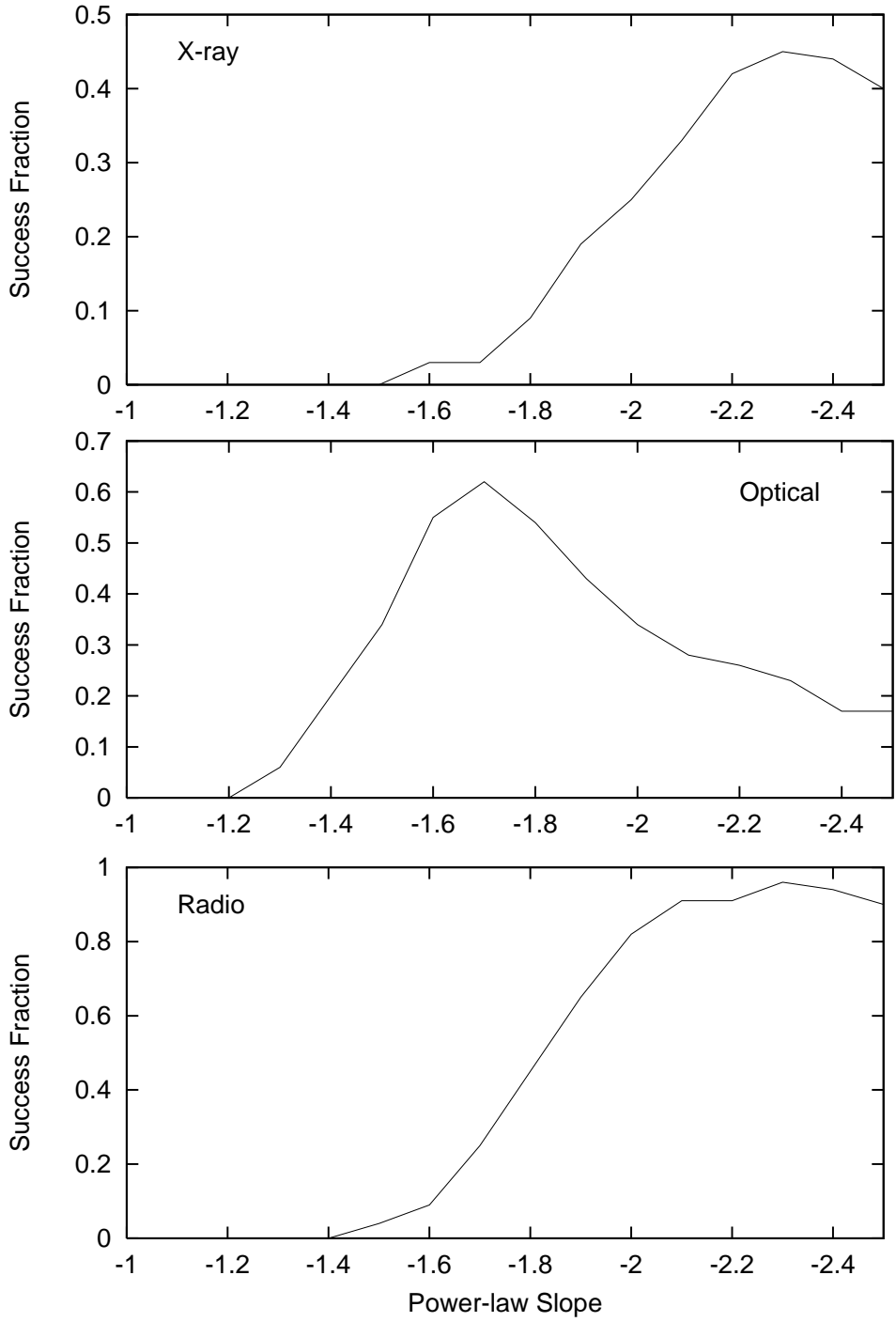}
\caption{Success fraction vs. slope for all three PSDs. The success fractions indicate the goodness of fit obtained from the PSRESP method (see text).}
\label{psdprob}
\end{figure}

The PSDs of the blazar 3C~279 at X-ray, optical, and radio frequencies show red noise behavior, i.e., there is higher amplitude variability on longer than on shorter timescales. The X-ray PSD is best fit with a simple power law of slope $-2.3\pm0.3$, for which the success fraction is 45\%. The slope of the optical PSD is $-1.7\pm0.3$ with success fraction 62\%, and for the radio PSD it is $-2.3\pm0.5$ with success fraction 96\%. The observed PSDs and their best-fit models are shown in Figure~\ref{psd}. The error bars on the slope represent the FWHM of the success fraction vs. slope curve (Figure~\ref{psdprob}). The rejection confidence, equal to one minus the success fraction, is much less than 90\% in all three cases (55\%, 38\%, and 4\% in the X-ray, optical, and radio wavebands, respectively). This implies that a simple power-law model provides an acceptable fit to the PSD at all three wavebands.

We also fit a broken power-law model to the X-ray PSD, setting the low-frequency slope at $-1.0$ and allowing the break frequency and the slope above the break over a wide range of parameters ($10^{-9}$ to $10^{-6}$ Hz and $-1.0$ to $-2.5$, respectively) while calculating the success fractions (\citet{mch06}). Although this gives lower success fractions than the simple power-law model for the whole paramater space, a break at a frequency $\lesssim 10^{-8}$ Hz with a high frequency slope as steep as $-2.4$ cannot be rejected at the 95\% confidence level.

\section{Cross-correlation Analysis and Results}
We employ the discrete cross-correlation function \citep[DCCF;][]{ede88} method to find the correlation between variations at pairs of wavebands. We bin the light curves of all three wavebands in 1-day intervals before performing the cross-correlation. In order to determine the significance of the correlation, we perform the following steps :\\
1. Simulation of $M$ (we use $M$=100) artificial light curves generated with a Monte-Carlo algorithm based on the shape and slope of the PSD as determined using PSRESP for both wavebands (total of 2$M$ light curves).\\
2. Resampling of the light curves with the observed sampling function.\\
3. Correlation of random pairs of simulated light curves (one at each waveband).\\
4. Identification of the peak in each of the $M$ random correlations. \\
5. Comparison of the peak values from step 4 with the peak value of the real correlation between the observed light curves. For example, if 10 out of 100 random peak values are greater than the maximum of the real correlation, we conclude that there is a 10\% chance of finding the observed correlation by chance. Therefore, if this percentage is low, then the observed correlation is significant even if the correlation coefficient is substantially lower than unity. 

\begin{figure}
\epsscale{.80}
\plotone{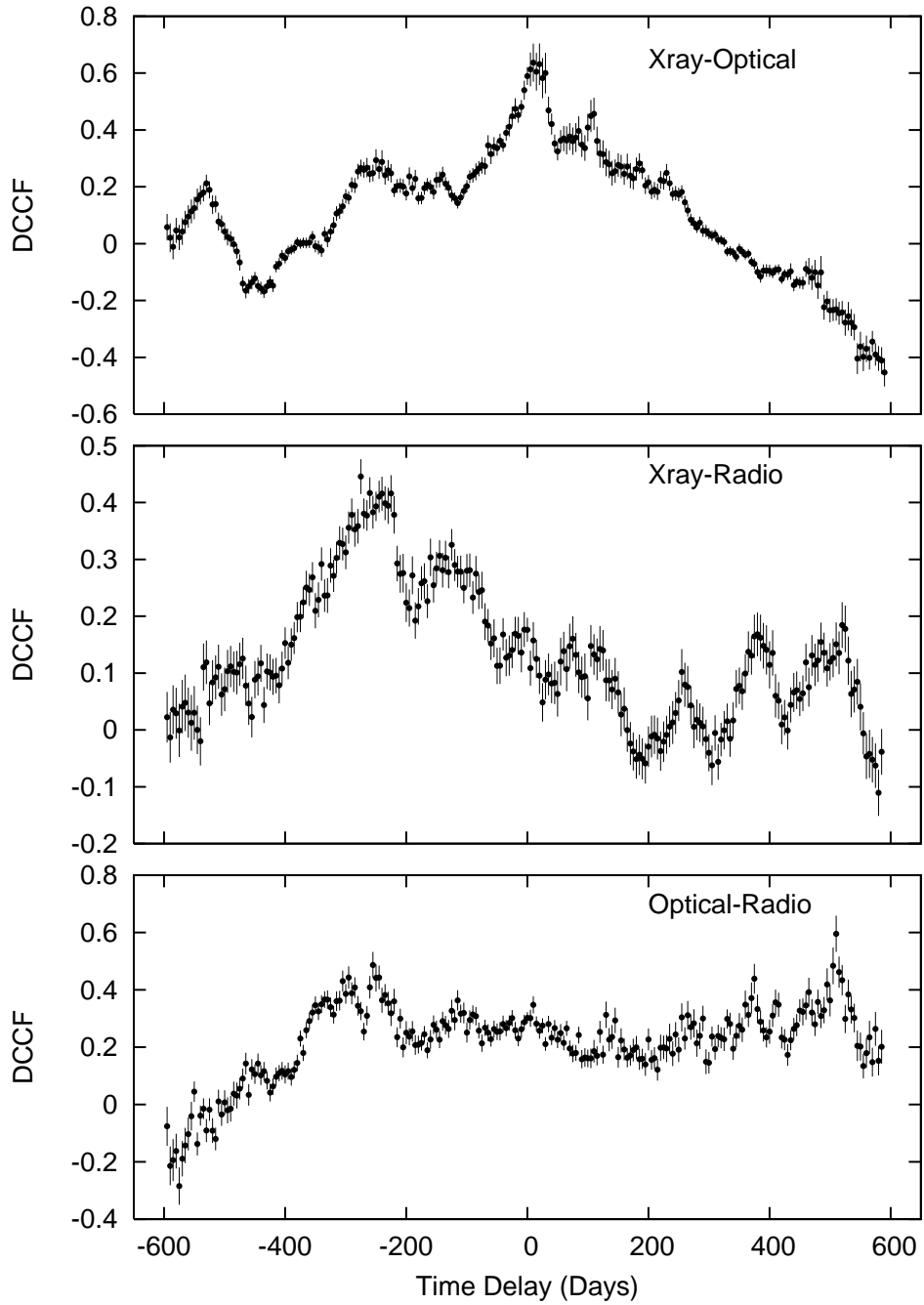}
\caption{Discrete cross-correlation function (DCCF) of the optical, X-ray, and radio monitor data. The time delay is defined as positive if the variations at the higher frequency waveband lag those at the lower frequency.}
\label{xopradcor}
\end{figure}

As determined by the DCCF (Figure~\ref{xopradcor}), the X-ray variations are correlated with those at both optical and radio wavelengths in 3C~279. The peak X-ray vs. optical DCCF is 0.66, which corresponds to a 98\% significance level. The peak X-ray vs. radio DCCF is relatively modest (0.42), with a significance level of 79\%. The radio-optical DCCF has a similar peak value (0.45) at a 62\% significance level. The cross-correlation also indicates that the optical variations lead the X-ray by $20\pm15$ and the radio by $260^{+60}_{-110}$ days, while X-rays lead the radio by $240^{+50}_{-100}$ days. The uncertainties in the time delays are the FWHM of the peaks in the correlation function. 

\subsection{Variation of X-ray/optical Correlation with Time} 
\begin{figure}
\epsscale{0.8}
\plotone{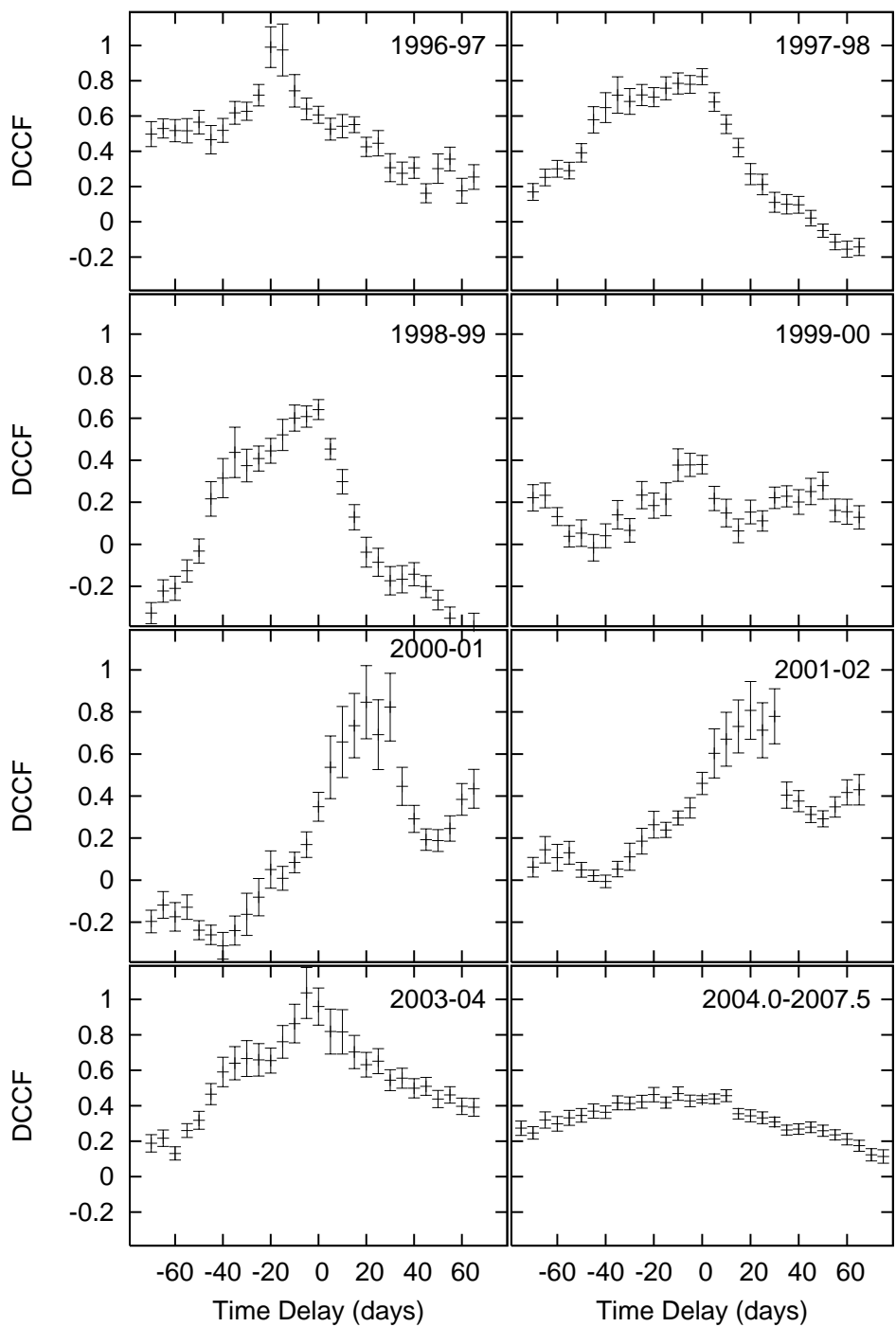}
\caption{Variation of X-ray/optical time lag across overlapping 2 intervals from the beginning of the first year to the end of the second, except for the bottom right panel, for which the interval is indicated more precisely. Notice the big change between 1998-99 and 2000-2001, when the time delay went from X-ray leading optical to the opposite sense. There is a major change between 2001-02 and 2003-04 as well, when the time delay went from optical leading X-ray to the opposite sense. In the last 4 years (bottom right panel) the correlation became weaker but maintained the negative time delay.}
\label{tw1}
\end{figure}

The X-ray and the optical light curves are correlated at a very high significance level. However, the uncertainty in the X-ray-optical time delay is comparable to the delay itself. To characterize the variation of the X-ray/optical time lag over the years, we divide both light curves into overlapping two-year intervals, and repeat the DCCF analysis on each segment. The result indicates that the correlation function varies significantly with time (Fig.~\ref{tw1}) over the 11 years of observation. Of special note are the following trends:\\
1. During the first four years of our program (96-97, 97-98, 98-99) the X-ray variations lead the optical (negative time lag). \\
2. There is a short interval of weak correlation in 1999-2000.\\
3. In 2000-01, the time delay shifts such that the optical leads the X-ray variations (positive time lag). This continues into the next interval (2001-02).\\ 
4. In 2002-03, there is another short interval of weak correlation (not shown in the figure).\\
5. In the next interval (2003-04), the delay shifts again to almost zero.\\
6. Over the next 3 years the correlation is relatively weak and the peak is very broad, centered at a slightly negative value.\\
This change of time lag over the years is the main reason why the peak value of the overall DCCF is significantly lower than unity. We discuss the physical cause of the shifts in cross-frequency time delay in $\S$5.2. 
 
\subsection{Correlation of X-ray Flux and Position Angle of the Inner Jet}
\begin{figure}
\epsscale{.80}
\plotone{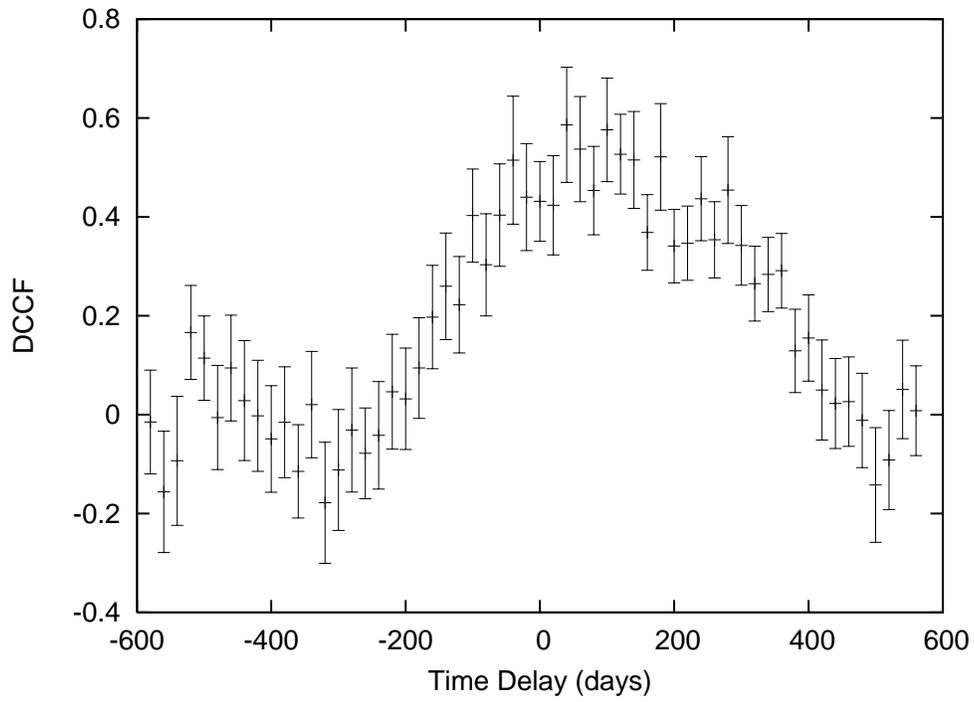}
\caption{Cross-correlation function of the X-ray light curve and the position angle of the jet. Changes in the position angle lead those in the X-ray flux by $80\pm 150$ days.}
\label{xposangle}
\end{figure}

We find a significant correlation (maximum DCCF=0.6) between the PA of the jet and the X-ray flux (see Figure~\ref{xposangle}). The changes in the position angle lead those in the X-ray flux by $80\pm150$ days. The large uncertainty in the time delay results from the broad, nearly flat peak in the DCCF. This implies that the jet direction modulates rather gradual changes in the X-ray flux instead of causing specific flares. This is as expected if the main consequence of a swing in jet direction is an increase or decrease in the Doppler beaming factor on a time scale of one or more years.

\subsection{Comparison of X-ray and Optical Flares}

We follow \citet{val99} by decomposing the X-ray and optical light curves into individual flares, each with exponential rise and decay. Our goal is to compare the properties of the major long-term flares present in the X-ray and optical lightcurves. 
Before the decomposition, we smooth the light curve using a Gaussian function with a 10-day FWHM smoothing time. 
We proceed by first fitting the highest peak in the smoothed light curve to an exponential rise and fall, and then subtracting the flare thus fit from the light curve. We do the same to the ``reduced" light curve, i.e., we fit the next highest peak. This reduces confusion created by a flare already rising before the decay of the previous flare is complete. We fit the entire light curve in this manner with a number of individual (sometimes overlapping) flares, leaving a residual flux much lower than the original flux at all epochs. We have determined the minimum number of flares required to adequately model the lightcurve to be 19 (X-ray) and 20 (optical), i.e., using more than 19 flares to model the X-ray light curve does not change the residual flux significantly. 

\begin{figure}
\epsscale{.80}
\plotone{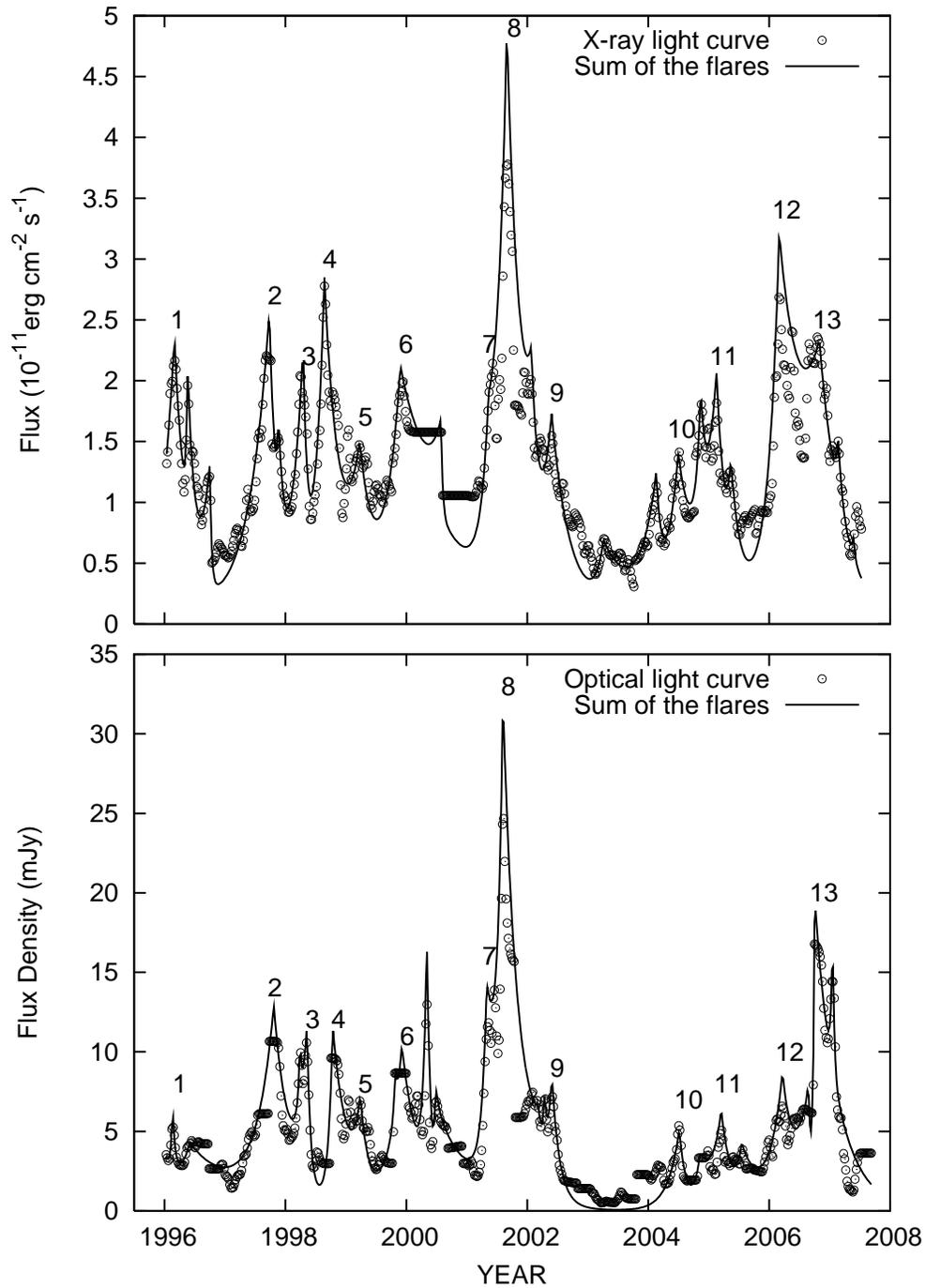}
\caption{Smoothed X-ray and optical light curves. Curves correspond to summed flux after modeling the light curve as a superposition of many individual flares. Thick horizontal strips in the X-ray light curve in 2000 correspond to epochs when no data are available. Flare pairs listed in Table 4 are marked with the respective ID numbers.}
\label{modelfit}
\end{figure}

Figure~\ref{modelfit} compares the smoothed light curves with the summed flux (sum of contributions from all the model flares at all epochs). We identify 13 X-ray/optical flare pairs in which the flux at both wavebands peaks at the same time within $\pm 50$ days. Since both light curves are longer than 4200 days and there are only about 20 significant flares during this time, it is highly probable that each of these X-ray/optical flare pairs corresponds to the same physical event. There are some X-ray and optical flares with no significant counterpart at the other waveband. We note that this does not imply complete absence of flaring activity at the other wavelength, rather that the corresponding increase of flux was not large enough to be detected in our decomposition of the smoothed light curve.

We calculate the area under the curve for each flare to represent the total energy output of the outburst. In doing so, we multiply the R-band flux density by the central frequency ($4.7\times10^{14}$ Hz) to estimate the integrated optical flux. For each of the flares, we determine the time of the peak, width (defined as the mean of the rise and decay times), and area under the curve from the best fit model. Table~\ref{xopcomp} lists the parameters of each flare pair, along with the ratio $\zeta_{\rm XO}$ of X-ray to optical energy output. The time delays of the flare pairs can be divided into three different classes: X-ray significantly leading the optical peak (XO, 6 out of 13), optical leading the X-ray (OX, 3 out of 13), and nearly coincident (by $< 10$ days, the smoothing length) X-ray and optical maxima (C, 4 out of 13). The number of events of each delay classification is consistent with the correlation analysis (Figure~\ref{tw1}). XO flares dominate during the first and last segments of our program, but OX flares occur in the middle. In both the DCCF and flare analysis, there are some cases just before and after the transition in 2001 when variations in the two wavebands are almost coincident (C flares). 

The value of $\zeta_{\rm XO}$ $\approx 1$ in 5 out of 13 cases; in one flare pair $\zeta_{\rm XO}=1.4$. In all the other cases it is less than unity by a factor of a few. In all the C flares $\zeta_{\rm XO}$ $\approx 1$, while in the 3 OX cases the ratio $\ll 1$. In the C pairs, the width of the X-ray flare profile $\sim 2$ times that of the optical, but in the other events the X-ray and optical widths are comparable.

\subsection{Flare-Ejection Correlation}
\begin{figure}
\epsscale{.80}
\plotone{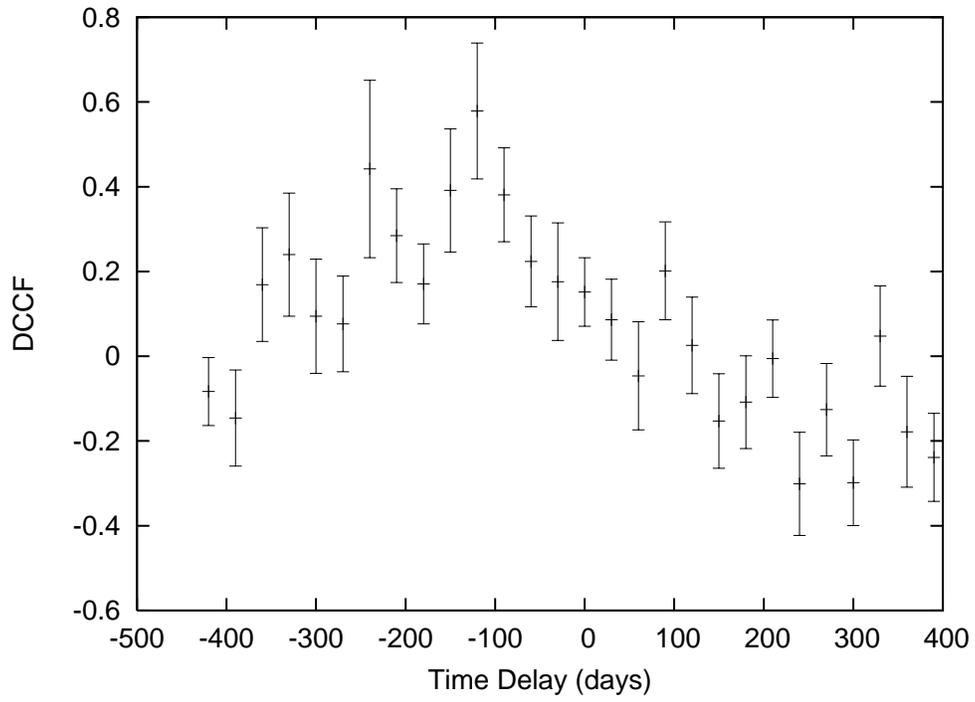}
\caption{Cross-correlation of the X-ray and 43 GHz core light curves. Changes in the X-ray flux lead those in the radio core by $130^{+70}_{-45}$ days.}
\label{xcore}
\end{figure}

The core region on VLBI images becomes brighter as a new superluminal knot passes through it \citep{sav02}; hence, maxima in the 43 GHz light curve of the core indicate the times of ejection of knots. We find that the core (Figure~\ref{corelc}) and X-ray light curves are well correlated (correlation coefficient of 0.6), with changes in the X-ray flux leading those in the radio core by $130^{+70}_{-45}$ days (see Figure~\ref{xcore}). The broad peak in the cross-correlation function suggests that the flare-ejection time delay varies over a rather broad range. This result is consistent with the finding of \citet{lin06} that high-energy flares generally occur during the rising portion of the 37 GHz light curve of 3C~279.

Since some flares can be missed owing to overlapping declines and rises of successive events, we can determine the times of superluminal ejections more robustly from the VLBA data. Table~\ref{ejecflare} lists the ejection times and apparent speeds of the knots identified by our procedure. We cannot, however, associate an X-ray flare with a particular superluminal ejection without further information, since flux peaks and ejection times are disparate quantities. We pursue this in a separate paper that uses light curves at five frequencies between 14.5 and 350 GHz to analyze the relationship between superluminal knots and flares in 3C~279.
 
\section{Discussion}
\subsection{Red Noise Behavior and Absence of a Break in the PSD}
The PSDs at all three wavebands are best fit with a simple power law which corresponds to red noise. The red noise nature---greater amplitudes on shorter time scales---of the flux variations at all three wavebands revealed by the PSD analysis is also evident from visual inspection of the light curves of 3C~279 (Figures~\ref{xdata}, \ref{opdata}, and \ref{raddata}). 

The PSD break frequency in BHXRBs and Seyferts scales with the mass of the black hole \citep{utt02, mch04, mch06, mar03, ede99}. Using the best-fit values and uncertainties in the relation between break timescale, black-hole mass, and accretion rate obtained by \citet{mch06}, we estimate the expected value of the break frequency in the X-ray PSD of 3C~279 to be $10^{-7.6\pm0.7}$ Hz, which is just within our derived lower limit of $10^{-8.5}$ Hz. Here we use a black-hole mass of $10^9$ M$_\sun$ \citep{woo02,liu06} for 3C~279 and a bolometric luminosity of the big blue bump of $4\times 10^{45}$ ergs/s \citep{har96}. If we follow \citet{mch06} and set the low-frequency slope of the X-ray PSD at $-1.0$ and allow the high-frequency slope to be as steep as $-2.4$, a break at a frequency $\lesssim 10^{-8}$ Hz cannot be rejected at the 95\% confidence level. An even longer light curve is needed to place more stringent limits on the the presence of a break at the expected frequency.

\subsection{Correlation between Light Curves at Different Wavebands}
The cross-frequency time delays uncovered by our DCCF analysis relate to the relative locations of the emission regions at the different wavebands, which in turn depend on the physics of the jet and the high-energy radiation mechanism(s). If the X-rays are synchrotron self-Compton (SSC) in nature, their variations may lag the optical flux changes owing to the travel time of the seed photons before they are up-scattered. As discussed in \citet{sok04}, this is an important effect provided that the angle ($\theta_{\rm obs}$) between the jet axis and the line of sight in the observer's frame is sufficiently small, $\lesssim 1.2\degr\pm0.2\degr$, in the case of 3C~279, where we have adopted the bulk Lorentz factor ($\Gamma = 15.5\pm2.5)$ obtained by \citet{jor05}. According to \citet{sok04}, if the emission region is thicker (thickness $\sim$ radius), the allowed angle increases to $2\degr\pm0.4\degr$. X-rays produced by inverse Compton scattering of seed photons from outside the jet (external Compton, or EC, process) may lag the low frequency emission for any value of $\theta_{\rm obs}$ between $0\degr$ and $90\degr$. However, in this case we expect to see a positive X-ray spectral index over a significant portion of a flare if $\theta_{\rm obs}$ is small, and the flares should be asymmetric, with much slower decay than rise \citep{sok05}. This is because the electrons that up-scatter external photons (radiation from a dusty torus, broad emission-line clouds, or accretion disk) to X-rays have relatively low energies, and therefore have long radiative cooling times. Expansion cooling quenches such flares quite slowly, since the EC flux depends on the total number of radiating electrons (rather than on the number density), which is relatively weakly dependent on the size of the emitting region.
 
Time delays may also be produced by frequency stratification in the jet. This occurs when the electrons are energized along a surface (e.g., a shock front) and then move away at a speed close to $c$ as they lose energy via synchrotron and IC processes \citep{mar85}. This causes the optical emission to be radiated from the region immediately behind the surface, with the IR emission arising from a somewhat thicker region and the radio from an even more extended volume. An optical to radio synchrotron flare then begins simultaneously (if opacity effects are negligible), but the higher-frequency peaks occur earlier. On the other hand, SSC (and EC) X-rays are produced by electrons having a range of energies that are mostly lower than those required to produce optical synchrotron emission \citep{mch99}. Hence, X-rays are produced in a larger region than is the case for the optical emission, so that optical flares are quenched faster and peak earlier. Flatter PSD in the optical waveband is consistent with this picture.

In each of the above cases, the optical variations lead those at X-ray energies. But in majority of the observed flares, the reverse is true. This may be explained by another scenario, mentioned by \citet{bot07}, in which the acceleration time scale of the highest-energy electrons is significantly longer than that of the lower-energy electrons, and also longer than the travel time of the seed photons and/or time lags due to frequency stratification. In this case, X-ray flares can start earlier than the corresponding optical events. 

We thus have a working hypothesis that XO (X-ray leading) flares are governed by gradual particle acceleration. OX events can result from either (1) light-travel delays, since the value of $\theta_{\rm obs}$ determined by \citet{jor05} ($2.1\degr\pm 1.1\degr$) is close to the required range and could have been smaller in 2001-03 when OX flares were prevalent, or (2) frequency stratification. One way to test this further would be to add light curves at $\gamma$-ray energies, as will be possible with the upcoming {\it Gamma Ray Large Area Space Telescope} (GLAST). If the X and $\gamma$ rays are produced by the same mechanism and the X-ray/optical time lag is due to light travel time, we expect the $\gamma$-ray/optical time lag to be similar. If, on the other hand, the latter delay is caused by frequency stratification, then it will be shorter, since IC $\gamma$ rays and optical synchrotron radiation are produced by electrons of similarly high energies.

If the synchrotron flare and the resultant SSC flare are produced by a temporary increase in the Doppler factor of the jet (due to a change in the direction, Lorentz factor, or both), then the variations in flux should be simultaneous at all optically thin wavebands. It is possible for the C flare pairs to be produced in this way. Alternatively, the C events could occur in locations where the size or geometry are such that the time delays from light travel and frequency stratification are very short compared to the durations of the flares.

As is discussed in {\S}4.2, the X-ray/optical time delay varies significantly over the observed period (see Table~\ref{xopcomp} and Fig.~\ref{tw1}). As we discuss above, XO flares can be explained by gradual acceleration of electrons. The switch in the time delay from XO to OX at the onset of the highest amplitude multi-waveband outburst, which occurred in 2001 (MJD 2000 to 2200), might then have resulted from a change in the jet that shortened the acceleration time significantly. This could have caused the time delay from light travel time of the SSC seed photons and/or the effects of gradients in maximum electron energy to become a more significant factor than the acceleration time in limiting the speed at which flares developed during the outburst period. However, the flux from the 43 GHz core also reached its maximum value in early 2002, and the apparent speed of the jet decreased from $\sim 17c$ in 2000 to $\sim 4-7c$ in 2001-2003. This coincided with the onset of a swing toward a more southerly direction of the trajectories of new superluminal radio knots. We hypothesize that the change in direction also reduced the angle between the jet and the line of sight, so that the Doppler factor $\delta$ of the jet increased significantly, causing the elevated flux levels and setting up the conditions for major flares to occur at all wavebands. The pronounced variations in flux during the 2001-2002 outburst cannot be explained solely by fluctuations in $\delta$, however, since the time delay switched to OX rather than C. Instead, a longer-term switch to a smaller viewing angle would have allowed the SSC light-travel delay to become important, causing the switch from XO to OX flares.

\subsection{Quantitative Comparison of X-ray and Optical Flares}
The relative amplitude of synchrotron and IC flares depends on which physical parameters of the jet control the flares.  The synchrotron flux is determined by the magnetic field $B$, the total number of emitting electrons $N_{\rm e}$, and the Doppler factor of the flow $\delta$.
The IC emission depends on the density of seed photons, number of electrons available for scattering $N_{\rm e}$, and $\delta$. An increase solely in $N_{\rm e}$ would enhance the synchrotron and EC flux by the same factor, while the corresponding SSC flare would have a higher relative amplitude owing to the increase in both the density of seed photons and number of scattering electrons. If the synchrotron flare were due solely to an increase in $B$, the SSC flare would have a relative amplitude similar to that of the synchrotron flare, since $B$ affects the SSC output only by increasing the density of seed photons. In this case, there would be no EC flare at all. Finally, if the synchrotron flare were caused solely by an increase in $\delta$, the synchrotron and SSC flux would rise by a similar factor, while the EC flare would be more pronounced, since the density and energy of the incoming photons in the plasma frame would both increase by a factor $\delta.$ Table~\ref{theory} summarizes these considerations. 

The location of the emission region should also have an effect on the multi-waveband nature of the flares. The magnetic field and electron energy density parameter $N_0$ both decrease with distance $r$ from the base of the jet: $B\sim r^{-b}$ and $N_0\sim r^{-n}$; we adopt $n=2$ and $b=1$ and assume a conical geometry. The cross-sectional radius $R$ of the jet expands with $r$, $R \propto r$. We have performed a theoretical calculation of the energy output of flares that includes the dependence on the location of the emission region. We use a computer code that calculates the synchrotron and SSC radiation from a source with a power-law energy distribution of electrons $N(\gamma)=N_0\gamma^{-s}$ within a range $\gamma_{\rm min}$ to $\gamma_{\rm max}$, where $\gamma$ is the energy in units of the electron rest mass. We introduce time variability of the radiation with an exponential rise and decay in $B$ and/or $N_0$ with time. In addition, we increase $\gamma_{max}$ with time in some computations in order to simulate the gradual acceleration of the electrons. The synchrotron emission coefficient is given by\\
\begin{equation}
j_\nu(\nu) = \frac{\nu N_0}{k^\prime}\int_{\gamma_{min}}^{\gamma_{max}} \gamma^{-s}(1-\gamma kt)^{s-2}\, d\gamma\ \int_{\frac{\nu}{k^\prime\gamma^2}}^{+\infty} K_{\frac{5}{3}}(\xi)\, d\xi,
\end{equation}
where $K_{\frac{5}{3}}$ is the modified Bessel function of the second kind of order $\frac{5}{3}$. We adopt $s=2.5$, consistent with the mean X-ray spectral index. The critical frequency, near which most of the synchrotron luminosity occurs, is given by $\nu_c= k^\prime \gamma^2$, while the synchrotron energy loss rate is given by $dE/dt=-k\gamma^2$. Both $k$ and $k^\prime$ are functions of $B$ and are given by $k=1.3\times10^{-9}B^2$ and $k^\prime=4.2\times10^{6}B$. The inverse-Compton (SSC in this case) emission coefficient is given by\\
\begin{equation}
j_\nu^C = \int_{\nu}\int_{\gamma}\frac{\nu_f}{\nu_i}j_\nu(\nu_i)R\sigma(\epsilon_i,\epsilon_f,\gamma) N(\gamma)\,d\gamma d\nu_i,
\end{equation}
where we approximate that the emission/scattering region is uniform (i.e., we ignore frequency stratification) and spherical with radius $R$. The Compton cross-section $\sigma$ is a function of $\gamma$ as well as the incident ($\nu_i$) and scattered $(\nu_f)$ frequencies of the photons:
\begin{equation}
\sigma(\nu_i,\nu_f,\gamma)=\frac{3}{32}\sigma_T\frac{1}{\nu_i\gamma^2}[8+2x-x^2+4x \ln (\frac{x}{4})],
\end{equation}
where $x\equiv\nu_f/(\nu_i \gamma^2)$ and $\sigma_T$ is the Thompson cross-section.

\begin{figure}
\epsscale{.80}
\plotone{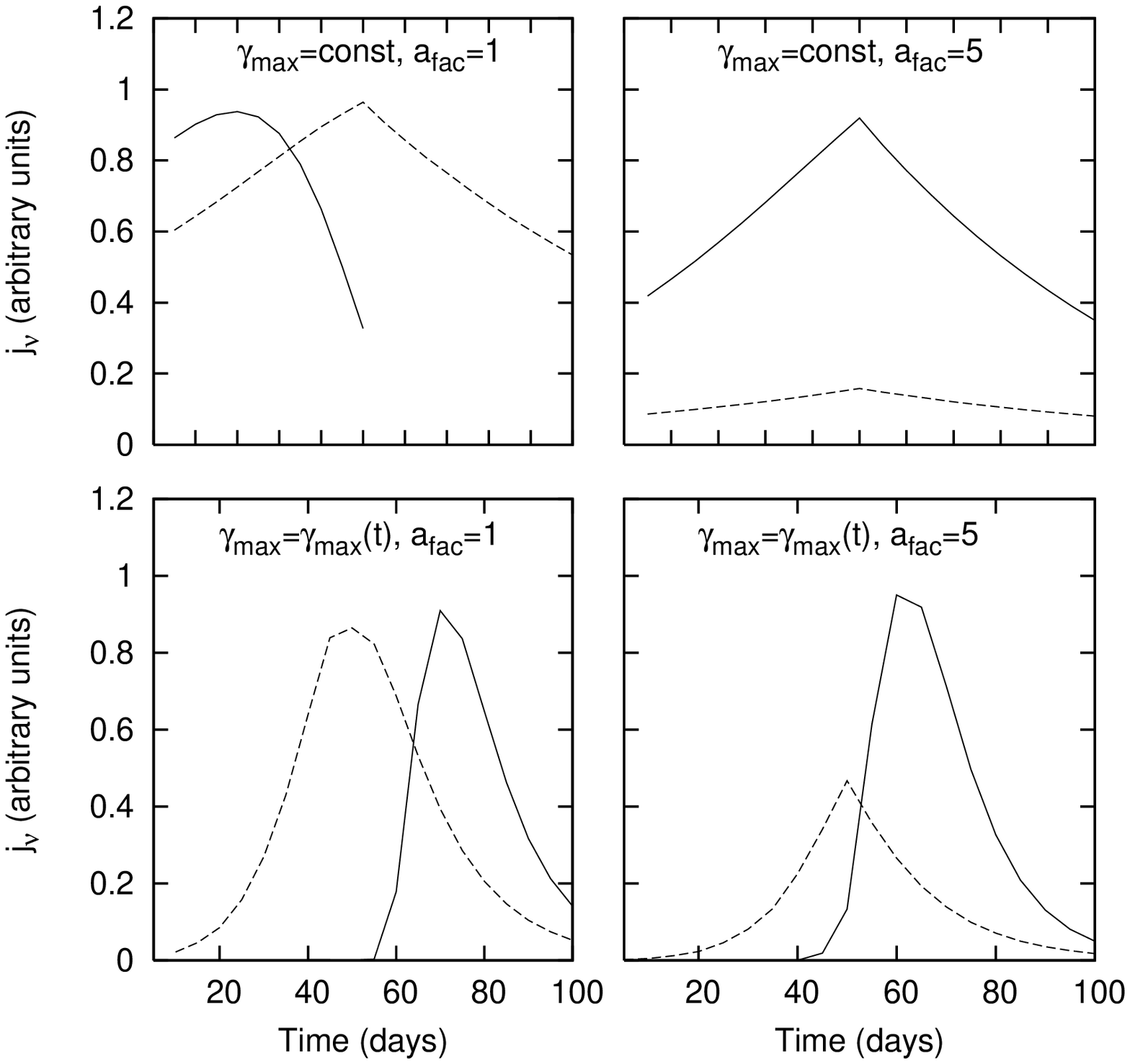}
\caption{Simulated synchrotron (solid curve) and SSC (dashed curve) flares. Here all flares are created by an exponential rise and decay in the magnetic field $B$ ($B$, $N_0$, and $R$ are functions of distance along the jet). In the bottom panels, $\gamma_{max}$ is increased linearly with time causing the SSC flux to peak ahead of the synchrotron flux. Flare amplitudes have been scaled such that they can be seen on the same plot. Normalization in the two upper panels is the same and that in the two lower panels is the same.}
\label{flcomp1}
\end{figure}

\begin{figure}
\epsscale{.80}
\plotone{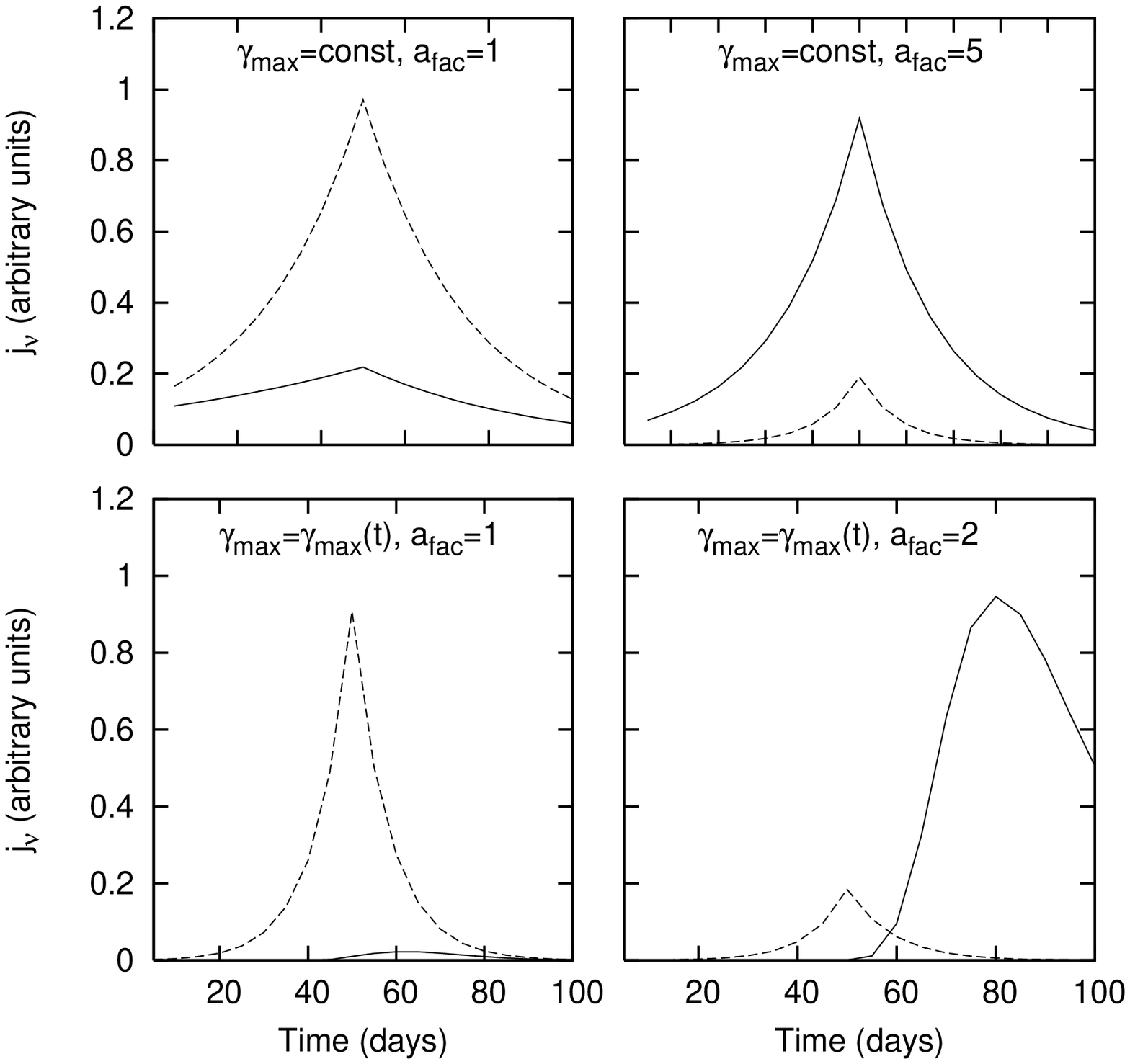}
\caption{Simulated synchrotron (solid curve) and SSC (dashed curve) flares. Here all flares are created by an exponential rise and decay in the magnetic field $N_0$ ($B$, $N_0$, and $R$ are functions of distance along the jet). In the bottom panels, $\gamma_{max}$ is increased linearly with time causing the SSC flux to peak ahead of the synchrotron flux. Flare amplitudes have been scaled such that they can be seen on the same plot. Normalization in the two upper panels is the same and that in the two lower panels is the same.}
\label{flcomp2}
\end{figure}

\begin{figure}
\epsscale{.80}
\plotone{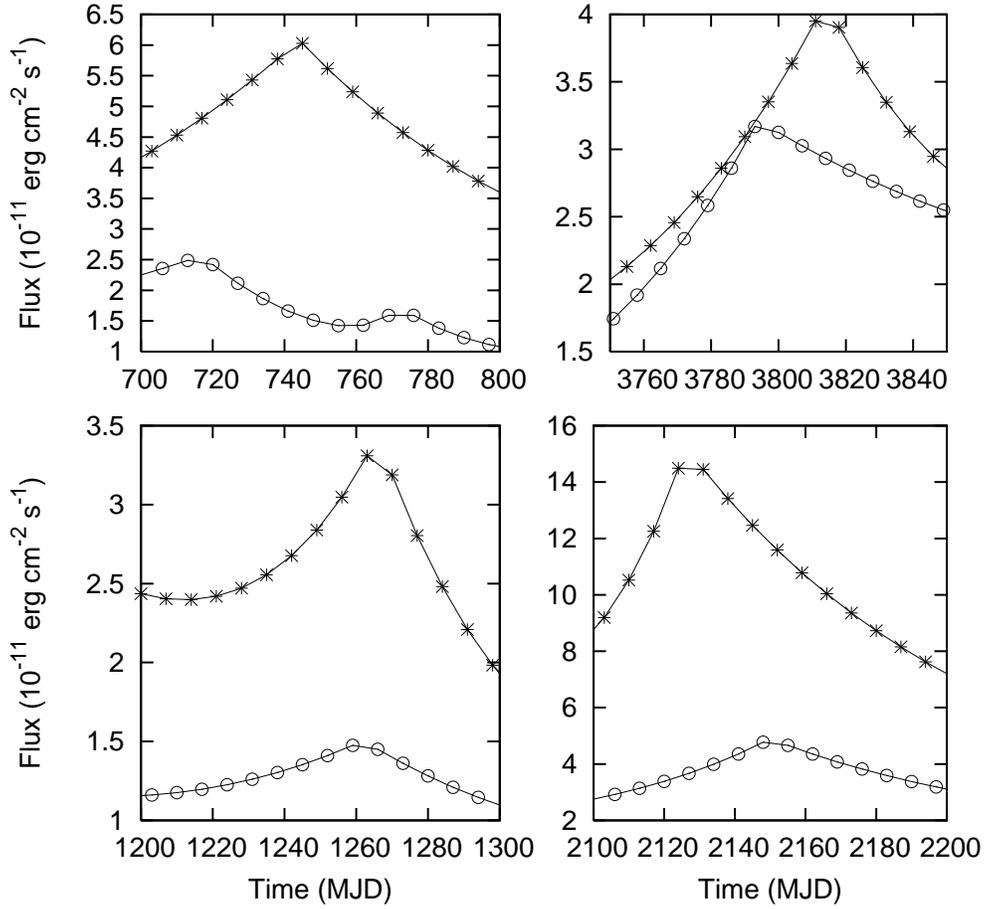}
\caption{Segments of real light curves (optical: asterisks, X-ray: open circles) over $\sim$ 100 day intervals similar to the length of the simulated light curves. Optical flux density is multiplied by R band central frequency ($4.7\times10^{14}$ Hz) to obtain the flux. The simulated light curves in Figures ~\ref{flcomp1} and ~\ref{flcomp2} have profiles similar to flares in the actual light curves.}
\label{flcompreal}
\end{figure}

In Figure~\ref{flcomp1}, the top left panel shows the synchrotron and SSC flares at a given distance $r$ from the base of the jet, where $r$ is a constant times a factor $a_{\rm fac}$ ($a_{\rm fac}$=1). The right panel shows the same at a farther distance ($a_{\rm fac}=5$). The flares are created by an exponential rise and decay in the $B$ field, while the value of $\gamma_{\rm max}$ is held constant. The total time of the flares is fixed at $10^{7}$ seconds (120 days). We can see from Figure~\ref{flcomp1} that at larger distances (right panel) the SSC flare has a lower amplitude than the synchrotron flare even though at smaller values of $r$ (left panel) they are comparable. The bottom panels of Figure~\ref{flcomp1} are the same as above except that the value of $\gamma_{max}$ increases with time, as inferred from the forward time delay that occurs in some events. We see the same effect as in the top panels. In Figure~\ref{flcomp2}, the top and bottom panels show similar results, but in these cases the flares are created by an exponential rise and fall in the value of $N_0.$ Again, the results are similar. Figure~\ref{flcompreal} shows segments of the actual X-ray and optical light curves, which are qualitatively similar to the simulated ones. The energy output of both synchrotron and SSC flares decreases with increasing $r$, but more rapidly in the latter. As a result, the ratio of SSC to synchrotron energy output $\zeta_{XO}$ decreases with $r$. The case $\zeta_{XO} \ll 1$ is therefore a natural consequence of gradients in $B$ and $N_0$. 

From Table~\ref{xopcomp}, we can see that in 7 out of 13 of the flare pairs $\zeta_{XO} \ll 1$. In one pair, $\zeta_{XO} >\;1$, and in all other pairs, $\zeta_{XO} \approx 1$. Our theoretical calculation suggests that the pairs where the ratio is less than 1 are produced at a larger distance from the base of the jet than those where the ratio $\gtrsim$ 1. The size of the emission region should be related to the cross-frequency time delay, since for all three explanations of the lag a larger physical size of the emission region should lead to a longer delay. We can then predict that the X-ray/optical time delay of the latter flares should be smaller than for the pairs with ratio $<\;1$. Indeed, inspection of Table~\ref{xopcomp} shows that, for most of the pairs, shorter time delays correspond to larger $\zeta_{XO}$, as expected. The smaller relative width of the optical C flares supports the conclusion that these occur closer to the base of the jet than the other flare pairs.

\section{Conclusions}
This paper presents well-sampled, decade-long light curves of 3C~279 between 1996 and 2007 at X-ray, optical, and radio wavebands, as well as monthly images obtained with the VLBA at 43 GHz. We have applied an algorithm based on a method by \citet{utt02} to obtain the broadband PSD of nonthermal radiation from the jet of 3C~279. Cross-correlation of the light curves allows us to infer the relationship of the emission across different wavebands, and we have determined the significance of the correlations with simulated light curves based on the PSDs. Analysis of the VLBA data yields the times of superluminal ejections and reveals time variations in the position angle of the jet near the core. We have identified 13 associated pairs of X-ray and optical flares by decomposing the light curves into individual flares. Comparison of the observed radiative energy output of contemporaneous X-ray and optical flares with theoretical expectations has provided a quantitative evaluation of synchrotron and SSC models. We have discussed the results by focusing on the implications regarding the location of the nonthermal radiation at different frequencies, physical processes in the jet, and the development of disturbances that cause outbursts of flux density in blazars. Our main conclusions are as follows: \\
(1) The X-ray, optical, and radio PSDs of 3C~279 are of red noise nature, i.e., there is higher amplitude variability at longer time scales than at shorter time scales. The PSDs can be described as power laws with no significant break, although a break in the X-ray PSD at a variational frequency $\lesssim 10^{-8}$ Hz cannot be excluded at a 95\% confidence level.\\
(2) X-ray variations correlate with those at optical and radio wavebands, as expected if nearly all of the X-rays are produced in the jet. The X-ray flux correlates with the projected jet direction, as expected if Doppler beaming modulates the mean X-ray flux level.\\ 
(3) X-ray flares are associated with superluminal knots, with the times of the latter indicated by increases in the flux of the core region in the 43 GHz VLBA images. The correlation has a broad peak at a time lag of $130^{+70}_{-45}$ days, with X-ray variations leading.\\
(4) Analysis of the X-ray and optical light curves and their interconnection indicates that the X-ray flares are produced by SSC scattering and the optical flares by the synchrotron process. Cases of X-ray leading the optical peaks can be explained by an increase in the time required to accelerate electrons to the high energies needed for optical synchrotron emission. Time lags in the opposite sense can result from either light-travel delays of the SSC seed photons or gradients in maximum electron energy behind the shock fronts. \\
(5) The switch to optical-leading flares during the major multi-frequency outburst of 2001 coincided with a decrease in the apparent speeds of knots from 16-17$c$ to 4-7$c$ and a swing toward the south of the projected direction of the jet near the core. This behavior, as well as the high amplitude of the outburst, can be explained if the redirection of the jet (only a 1$\degr$-2$\degr$ change is needed) caused it to point closer to the line of sight than was the case before and after the 2001-02 outburst. \\
(6) Contemporaneous X-ray and optical flares with similar radiative energy output originate closer to the base of the jet, where the cross-section of the jet is smaller, than do flares in which the optical energy output dominates. This is supported by the longer time delay in the latter case. This effect is caused by the lower electron density and magnetic field and larger cross-section of the jet as the distance from the base increases.

Further progress in our understanding of the physical structures and processes in compact relativistic jets can be made by increasing the number of wavebands subject to intense monitoring. Expansion of such monitoring to a wide range of $\gamma$-ray energies will soon be possible when GLAST scans the entire sky several times each day. When combined with similar data at lower frequencies as well as VLBI imaging, more stringent tests on models for the nonthermal emission from blazars will be possible.

\acknowledgments
We thank P. Uttley for many useful discussions. The research at Boston University was funded in part by the National Science Foundation (NSF) through grant AST-0406865 and by NASA through several RXTE Guest Investigator Program grants, most recently NNX06AG86G, and Astrophysical Data Analysis Program grant NNX08AJ64G. The University of Michigan Radio Astronomy Observatory was supported by funds from the NSF, NASA, and the University of Michigan. The VLBA is an instrument of the National Radio Astronomy Observatory, a facility of the National Science Foundation operated under cooperative agreement by Associated Universities, Inc.  The Liverpool Telescope is operated on the island of La Palma by Liverpool John Moores University in the Spanish Observatorio del Roque de los Muchachos of the Instituto de Astrofisica de Canarias, with financial support from the UK Science and Technology Facilities Council.

{\it Facilities:} VLBA, RXTE, Liverpool:2m, Perkins

\appendix
\section{Power Spectrum Response Method (PSRESP)}
In this study, we determine the shape and slope of the PSD of the light curves of 3C~279 using the PSRESP method \citep{utt02}. This involves the following steps:\\
1. Calculation of the PSD of the observed light curve (PSD$_{\rm obs})$ with formulas (1) and (2).\\
2. Simulation of $M$ artificial light curves of red noise nature with a trial shape (simple power law, broken power law, bending power law, etc.) and slope. We use $M$ = 100.\\
3. Resampling of the simulated light curves with the observed sampling function.\\
4. Calculation of the PSD of each of the resampled simulated light curves (PSD$_{\rm sim,i}$, i=1, $M$). The resampling with the observed sampling function (which is irregular) adds the same distortions to the simulated PSDs that are present in the real PSD $(\rm PSD_{\rm obs})$.\\
5. Calculation of two functions similar to $\chi^2$: \\
\begin{equation}
\chi^2_{\rm obs}=\sum_{\nu=\nu_{\rm min}}^{\nu_{\rm max}}{\frac{(PSD_{\rm obs}-\overline{\rm PSD}_{\rm sim})^2}{(\Delta PSD_{\rm sim})^2}} 
\end{equation}
and
\begin{equation}
\chi^2_{\rm dist,i}=\sum_{\nu=\nu_{\rm min}}^{\nu_{(\rm max}}{\frac{(PSD_{\rm sim,i}-\overline{\rm PSD}_{\rm sim})^2}{(\Delta PSD_{\rm sim})^2}}, 
\end{equation}
where $\overline{\rm PSD}_{\rm sim}$ is the average of $(\rm PSD_{\rm sim,i})$ and
$\Delta \rm PSD_{\rm sim}$ is the standard deviation of $(\rm PSD_{\rm sim,i})$, with i=1, $M$.\\
6. Comparison of $\chi^2_{\rm obs}$ with the $\chi^2_{\rm dist}$
distribution. Let $m$ be the number of $\chi^2_{\rm dist,i}$ for which $\chi^2_{\rm obs}$ is smaller than $\chi^2_{\rm dist,i}$ . Then $m$/$M$ is the success fraction of that trial shape and slope, a measure of its success at representing the shape and slope of the intrinsic PSD. \\
7. Repetition of the entire procedure (steps 2 to 6) for a set of trial shapes and slopes of the initial simulated PSD to determine the shape and slope that gives the highest success fraction. We scan a range of trial slopes from $-1.0$ to $-2.5$ in steps of $0.1$ for the simple power-law fit.

We perform a few additional steps to overcome the distorting effects of finite length and discontinuous sampling of the light curves. These steps are implicitly included in the light curve simulation (step 2). The light curve of an astronomical source is essentially infinitely long, but we have sampled a 10-year long interval of it and are calculating the PSD based on that interval. As a result, power from longer (than observed) time scales ``leaks" into the shorter time scales and hence distorts the observed PSD. This effect, called ``red noise leak" (RNL), can be accounted for in PSRESP. We overcome this by simulating light curves that are more than 100 times longer than the observed light curve. As a consequence, the resampled simulated light curves are a small subset of the originally simulated ones, and similar RNL distortions are included in $\rm PSD_{\rm sim,i}$ that are present in $\rm PSD_{\rm obs}$. 

On the other hand, if a light curve is not continuously sampled, power from frequencies higher than the Nyquist frequency ($\nu_{\rm Nyq}$) is shifted or ``aliased" to frequencies below $\nu_{\rm Nyq}$. The observed PSD in that case will be distorted by the aliased power, which is added to the observed light curve from timescales as small as the exposure time ($T_{exp}$) of the observation (about 1000 seconds for the X-ray light curve). Ideally, we should account for this by simulating light curves with a time-resolution as small as 1000 seconds so that the same amount of aliasing occurs in the simulated data. This involves excessive computing time for decade-long light curves. To avoid this, we follow \citet{utt02} by simulating light curves with a resolution 10$T_{exp}$ . To calculate the aliasing power from timescales from $T_{exp}$  to 10$T_{exp}$, we use an analytic approximation of the level of power added to all frequencies by the aliasing, given by\\
\begin{equation} 
P_{C} = \frac{1}{\nu_{\rm Nyq} - \nu_{\rm min}}\int_{\nu_{\rm Nyq}}^{(2T_{\rm exp})^{-1}} P(\nu)d\nu.
\end{equation}
We use PSRESP to account for aliasing at frequencies lower than $(10T_{exp})^{-1}$.

We also add Poisson noise to the simulated light curves$\colon$ \\
\begin{equation}
P_{\rm noise}=\frac{\sum_{i=1}^{N}(\sigma(i))^2}{N(\nu_{\rm Nyq} - \nu_{\rm min})},
\end{equation}
where $\sigma(i)$ are observational uncertainties. For details of the noise processes, readers should refer to \citet{utt02}

The goal of adding the noise and resampling with the observed sampling function is to simulate a dataset that has the same properties, including the imperfections, as the observed one. This provides a physically meaningful comparison of the observed PSD with the distribution of the simulated PSDs.

\clearpage
\begin{table}
\begin{center}
\caption{Start and end times of observations presented in this study.\label{data}}
\begin{tabular}{ccccccc}\\
\tableline
&\multicolumn{2}{c}{X-ray} &\multicolumn{2}{c}{Optical}&\multicolumn{2}{c}{Radio}\\ \tableline
Dataset & Start  & End & Start  & End & Start  & End\\
\tableline\tableline
Longlook &Dec 96 &Jan 97 & Mar 05 & Jun 05  \\
Medium &Nov 03 &Sep 04 & Jan 04 & Jul 04& Mar 05 & Sep 05\\  
Monitor &Jan 96 &Jun 07 & Jan 96 & Jun 07& Jan 96 & Sep 07\\  
\tableline
\end{tabular}
\end{center}
\end{table}
\clearpage


\begin{deluxetable}{cccccccc}
\tablewidth{0pt}
\tablecaption{X-ray (2-10 keV) light curve data from RXTE (first 5 rows). \label{xdatatable}}
\tablehead{
\colhead{MJD\tablenotemark{1}} & \colhead{Exp. (s)} & \colhead{Flux\tablenotemark{2}} & \colhead{Error} & \colhead{$\alpha$\tablenotemark{3}} & \colhead{Err} & \colhead{count rate\tablenotemark{4}} & \colhead{Err}}
\startdata
 105.1275 &  688 & 1.314e-11 & 1.818e-12 & 0.466 & 0.226 & 2.766e+00 & 1.722e-01 \\
 106.2820 &  720 & 1.343e-11 & 1.648e-12 & 0.489 & 0.206 & 2.810e+00 & 1.602e-01 \\
 107.2621 &  672 & 1.150e-11 & 1.829e-12 & 0.425 & 0.246 & 2.393e+00 & 1.625e-01 \\
 108.2144 &  672 & 1.299e-11 & 1.568e-12 & 0.544 & 0.221 & 2.747e+00 & 1.670e-01 \\
 109.2619 &  624 & 1.173e-11 & 1.440e-12 & 0.597 & 0.245 & 2.474e+00 & 1.688e-01 \\
 ...      & ...  & ...       &    ...    &   ... & ...   & ...       &  ...      \\  
\enddata
\tablenotetext{1}{MJD = Julian date minus 2450000}
\tablenotetext{2}{Units: erg cm$^{-2}$ s$^{-1}$}
\tablenotetext{3}{``Energy'' spectral index}
\tablenotetext{4}{Photon counts s$^{-1}$ per PCU detector}
\end{deluxetable}

\begin{deluxetable}{ccccc}
\tablewidth{0pt}
\tablecaption{Optical (R Band) light curve data (first 5 rows). \label{opdatatable}}
\tablehead{
\colhead{MJD\tablenotemark{1}} & \colhead{mag} & \colhead{err} & \colhead{Flux Density\tablenotemark{2}} & \colhead{err}}
\startdata
  46.9514  &  14.383  &  0.007  &   5.837  &  0.038  \\
  71.7330  &  15.170  &  0.032  &   2.827  &  0.082  \\
  86.6356  &  15.386  &  0.032  &   2.317  &  0.067  \\
  91.6813  &  14.740  &  0.032  &   4.201  &  0.122  \\
 100.6226  &  14.878  &  0.032  &   3.700  &  0.107  \\
 ...       & ...      & ...     &    ...   &   ...   \\         
\enddata
\tablenotetext{1}{MJD = Julian date minus 2450000}
\tablenotetext{2}{Units: mJy}
\end{deluxetable}

\begin{deluxetable}{ccc}
\tablewidth{0pt}
\tablecaption{Radio (14.5 GHz) light curve data (first 5 rows). \label{raddatatable}}
\tablehead{
\colhead{MJD\tablenotemark{1}} & \colhead{Flux Density (Jy)} & \colhead{error}}
\startdata
  97.00  &  17.51  &  0.20  \\
  98.00  &  17.74  &  0.19  \\
 101.00  &  18.62  &  0.15  \\
 103.00  &  17.54  &  0.44  \\
 104.00  &  18.28  &  0.22  \\
 ...     & ...     & ...    \\  
\enddata
\tablenotetext{1}{MJD = Julian date minus 2450000}
\end{deluxetable}

\begin{table}
\begin{center}
\caption{Ejection times, apparent speeds, and position angle of superluminal knots.\label{ejecflare}}
\begin{tabular}{cccccccc}\\
\tableline
Knot & $T_0$& $T_0$ (MJD\tablenotemark{1}) & $\beta_{\rm app}$ & $\theta$ (deg)\tablenotemark{2} \\
\tableline
C8  &  1996.09$\pm$  0.10  &    115$\pm$ 36  &    5.4$\pm$ 0.7  &  $-130\pm$  3\\
C9  &  1996.89$\pm$  0.12  &    407$\pm$ 44  &   12.9$\pm$ 0.3  &  $-131\pm$  5\\
C10 &  1997.24$\pm$  0.16  &    536$\pm$ 58  &    9.9$\pm$ 0.5  &  $-132\pm$  6\\
C11 &  1997.59$\pm$  0.11  &    662$\pm$ 40  &   10.1$\pm$ 1.2  &  $-135\pm$  4\\
C12 &  1998.56$\pm$  0.09  &   1016$\pm$ 33  &   16.9$\pm$ 0.4  &  $-129\pm$  3\\
C13 &  1998.98$\pm$  0.07  &   1174$\pm$ 26  &   16.4$\pm$ 0.5  &  $-130\pm$  4\\
C14 &  1999.50$\pm$  0.09  &   1360$\pm$ 33  &   18.2$\pm$ 0.7  &  $-135\pm$  6\\
C15 &  1999.85$\pm$  0.05  &   1487$\pm$ 18  &   17.2$\pm$ 2.3  &  $-131\pm$  7\\
C16 &  2000.27$\pm$  0.05  &   1642$\pm$ 18  &   16.9$\pm$ 3.5  &  $-140\pm$  8\\
C17 &  2000.96$\pm$  0.12  &   1895$\pm$ 44  &    6.2$\pm$ 0.5  &  $-133\pm$ 12\\
C18 &  2001.40$\pm$  0.16  &   2054$\pm$ 58  &    4.4$\pm$ 0.7  &  $-150\pm$  8\\
C19 &  2002.97$\pm$  0.12  &   2648$\pm$ 44  &    6.6$\pm$ 0.6  &  $-133\pm$  7\\
C20 &  2003.39$\pm$  0.10  &   2781$\pm$ 44  &    6.0$\pm$ 0.5  &  $-155\pm$ 10\\
C21 &  2004.75$\pm$  0.05  &   3280$\pm$ 18  &   16.7$\pm$ 0.3  &  $-147\pm$  7\\
C22 &  2005.18$\pm$  0.06  &   3434$\pm$ 22  &   12.4$\pm$ 1.2  &  $-102\pm$17\\
C23 &  2006.41$\pm$  0.15  &   3888$\pm$ 55   &  16.5$\pm$ 2.3  &  $-114\pm$  5\\

\tableline 
\end{tabular}
\tablenotetext{1}{MJD = Julian date minus 2450000}
\tablenotetext{2}{Average position angle of knot within 1 mas of the core.}
\end{center}
\end{table}
\clearpage
\begin{table}
\begin{center}
\caption{Parameters of the light curves for calculation of PSD.\label{lcprop}}
\begin{tabular}{ccccccc}\\
\tableline
  &Dataset & T (days) & $\Delta$T (days) & log($f_{\rm min}$) & log($f_{\rm max}$) & N$_{\rm points}$ \\
\tableline
        & Longlook &55.0   &0.5  &-6.67  &-4.93& 111  \\
 X-ray  & Medium   &301.0  &5.0  &-7.40  &-5.93& 127  \\
        & Monitor  &4150.0 &25.0 &-8.55  &-6.63& 1213 \\ 
\tableline
         & Longlook &86.0   &1.0  &-6.86  &-5.23& 94 \\
 Optical & Medium   &185.0  &5.0  &-7.17  &-5.92& 77 \\
         & Monitor  &4225.0 &25.0 &-8.55  &-6.63& 995 \\ 
\tableline
 Radio  & Medium   &189.0  &4.0  &-7.18  &-5.83& 59\\
        & Monitor  &3984.0 &25.0 &-8.53  &-6.63& 609\\  
\tableline
\end{tabular}
\end{center}
\end{table}
\clearpage
\begin{table}
\begin{center}
\caption{Total energy output (area) and widths of flare pairs.\label{xopcomp}}
\begin{tabular}{cccccccccc}\\
\tableline
ID&\multicolumn{3}{c}{X-ray} &\multicolumn{3}{c}{Optical}&$\Delta T$ & TDC &$\zeta_{XO}$\tablenotemark{4}\\ 
&Time\tablenotemark{1}  & Area \tablenotemark{2}& Width\tablenotemark{3}& Time\tablenotemark{1}  & Area\tablenotemark{2}& Width\tablenotemark{3}& (days) \\
\tableline\tableline
1&   119 (1996.10)&   33.2&    6.0&   134 (1996.14) &  76.3&  12.0& -15 & XO & 0.44\\
2&   717 (1997.74)&  371.8&   90.0&   744 (1997.81) &  929.5& 97.5& -27 & XO & 0.40\\
3&   920 (1998.30)&  172.8&   50.0&   944 (1998.36) &  278.2& 35.0& -24 & XO & 0.62\\
4&  1050 (1998.65)&  338.7&   70.0&  1099 (1998.79) &  502.8& 57.5& -49 & XO & 0.67\\ 
5&  1263 (1999.24)&  160.7&   77.5&  1266 (1999.24) &  168.9& 42.5&  -3 & C  & 0.95\\
6&  1509 (1999.91)&  699.8&  202.5&  1517 (1999.93) &  715.4& 90.0&  -8 & C  & 0.98\\
7&  2045 (2001.38)&  103.7&   60.0&  2029 (2001.33) &  568.3& 65.0&  16 & OX & 0.18\\
8&  2151 (2001.67)&  453.6&   62.5&  2126 (2001.60) & 1188.3& 57.5&  25 & OX & 0.38\\
9&  2419 (2002.40)&  193.5&   80.0&  2422 (2002.41) &  222.6& 40.0&  -3 & C  & 0.87\\
10&  3185 (2004.50)&  191.8&   92.5&  3191 (2004.52) &  218.6& 55.0&  -6 & C  & 0.88\\
11&  3416 (2005.13)&  217.7&   70.0&  3444 (2005.21)&  198.7& 50.0& -28 & XO & 1.09\\
12&  3792 (2006.16)&  525.3&  120.0&  3814 (2006.22)&  375.6& 67.5& -22 & XO & 1.40\\
13&  4035 (2006.83)&  324.2&  105.0&  4008 (2006.76)& 1068.4& 75.0&  27 & OX & 0.30\\
\tableline 
\end{tabular}
\tablenotetext{1}{Units: MJD (yr)}
\tablenotetext{2}{Units: $10^{-6}$ erg cm$^{-2}$}
\tablenotetext{3}{Units: days}
\tablenotetext{4}{Ratio of X-ray to optical energy output integrated over flare}
\end{center}
\end{table}

\begin{table}
\tabletypesize{\scriptsize}
\begin{center}
\caption{Ratio of synchrotron to SSC or EC flare amplitude expected when a physical parameter varies.\label{theory}}
\begin{tabular}{ccc}
\\
\tableline\tableline
Parameter varied& SSC/Synch  & EC/Synch \\
\tableline
$\delta$ &$\approx 1$ & $>1$  \\
$N_{\rm e}$ &$>1$ &$\approx 1$    \\  
B &$\approx 1$  &$<1$   \\  
\tableline
\end{tabular}
\end{center}
\end{table}

\clearpage

\clearpage

\clearpage

\clearpage

\clearpage

\clearpage

\clearpage

\clearpage

\clearpage

\clearpage

\clearpage

\clearpage

\clearpage

\clearpage

\clearpage

\clearpage


\clearpage

\clearpage

\clearpage

\clearpage

\clearpage
\clearpage
\clearpage

\end{document}